\documentclass[12pt, draftclsnofoot,onecolumn, letterpaper]{IEEEtran}
\usepackage[T1]{fontenc}
\usepackage{color}
\usepackage{amsmath}
\usepackage{url}
\usepackage{array}
\usepackage{fancyhdr}
\usepackage{latexsym}
\usepackage{diagbox}
\usepackage{epsfig}
\usepackage{amssymb}
\usepackage{amsfonts}
\usepackage{amsxtra}
\usepackage{xspace}
\usepackage{makeidx}
\usepackage{graphics}
\usepackage{theorem}
\usepackage{subfigure}
\usepackage{algorithm}
\usepackage{algorithmic}
\usepackage{multirow}
\usepackage{cite}
\DeclareMathOperator*{\minimize}{minimize}
\DeclareMathOperator*{\maximize}{maximize}
\newtheorem{lemma}{\textbf{Lemma}}
\newtheorem{theorem}{\textbf{Theorem}}
\newtheorem{proposition}{\textbf{Proposition}}
\begin{document}

\title{Partially-Connected Hybrid Beamforming for Spectral Efficiency Maximization via a Weighted MMSE Equivalence}

\author{Xingyu~Zhao,~\IEEEmembership{Student Member,~IEEE,} Tian~Lin,~\IEEEmembership{Student Member,~IEEE,} Yu~Zhu,~\IEEEmembership{Member,~IEEE,} and Jun~Zhang,~\IEEEmembership{Senior Member,~IEEE}
\thanks{This work was supported by National Natural Science Foundation of China under Grant No. 61771147.}
\thanks{X. Zhao, T. Lin and Y. Zhu are with the Department of Communication Science and Engineering, Fudan University, Shanghai, China (e-mail: xingyuzhao19@fudan.edu.cn, lint17@fudan.edu.cn, zhuyu@fudan.edu.cn).}
\thanks{J. Zhang is with the Department of Electronic and Information Engineering, The Hong Kong Polytechnic University, Hong Kong (e-mail: jun-eie.zhang@polyu.edu.hk).}}

\maketitle
\begin{abstract}
Hybrid beamforming (HBF) is an attractive technology for practical massive multiple-input and multiple-output (MIMO) millimeter wave (mmWave) systems. Compared with the fully-connected HBF architecture, the partially-connected one can further reduce the hardware cost and power consumption. However, the special block diagonal structure of its analog beamforming matrix brings additional design challenges. In this paper, we develop effective HBF algorithms for spectral efficiency maximization (SEM) in mmWave massive MIMO  systems with the partially-connected architecture. One main contribution is that we prove the equivalence of the SEM problem and a matrix weighted sum mean square error  minimization (WMMSE) problem, which leads to a convenient algorithmic approach to directly tackle the SEM problem. Specifically, we decompose the equivalent WMMSE problem into the hybrid precoding and hybrid combining subproblems, for which both the optimal digital precoder and combiner have closed-form solutions. For the more challenging analog precoder and combiner, we propose an element iteration based algorithm and a manifold optimization based algorithm.  Finally, the hybrid precoder and combiner are alternatively updated. The overall HBF algorithms are proved to monotonously increase the spectral efficiency and converge. Furthermore, we also propose modified algorithms with reduced computational complexity and finite-resolution phase shifters. Simulation results demonstrate that the proposed HBF algorithms achieve significant performance gains over conventional algorithms.\\

\end{abstract}

\IEEEpeerreviewmaketitle

\section{Introduction}\label{sec:introduction}
The strong desire for supporting ultra-high-speed data transmission has promoted the investigation and application of millimeter wave (mmWave) communications due to its advantage of providing huge spectrum resources \cite{Heath:2016,Roh:2014,Akdeniz:2014,Andrews:2014,XUE:2014}. To overcome the severe path loss and penetration loss of the mmWave propagation channel while considering the stringent constraint of the mmWave hardware cost and power consumption, the combination of massive multiple-input and multiple-output (MIMO) and hybrid beamforming (HBF) has recently become an attractive technology \cite{Zhang:2020,Larsson:2014,Molisch:2017,Ayach:2014}. However, in the broadband scenario, the HBF design is quite challenging because of the joint optimization of a larger number of low-dimensional digital beamformers for all the subcarriers and a high-dimensional analog beamformer for the whole bandwidth with the specific constant modulus constraint due to the implementation of phase shifters \cite{Yu:2016,Sohrabi:2017,Lin:2019}. 

\subsection{Related Works and Motivations}\label{subsec:background}
Previous works have spent significant efforts on the fully-connected HBF architecture \cite{Yu:2016,Lin:2019,Ayach:2014,Sohrabi:2016,Cong:2018,Nguyen:2017,HAN:2017,Rusu:2016,Mirza:2017,Yu:2019}, where each radio frequency (RF) chain is connected to all the antennas. In \cite{Ayach:2014}, the authors regarded the HBF design as a matrix factorization problem by minimizing the Euclidean distance between the HBF matrix and the fully-digital beamforming matrix. The well-known orthogonal matching pursuit algorithm was applied to the HBF design with the motivation of exploiting the sparse characteristics of the mmWave propagation channel. However, the constraint that the analog beamforming matrix must be taken in a limited feasible set space led to certain performance loss. In \cite{Yu:2016}, a manifold optimization (MO) based algorithm has been proposed to directly deal with the constant modulus constraint for better performance. In \cite{Sohrabi:2016}, instead of solving the matrix factorization problem, the authors directly targeted the original SEM problem and proposed some iterative algorithms. In \cite{Lin:2019}, the minimum mean square error (MMSE) criterion has been taken for the HBF optimization, based on which the HBF design problem can be shown to be decomposed into two subproblems with respect to hybrid precoding and hybrid combining optimization and solved in a unified way.


Although the fully-connected architecture has the potential of achieving the full beamforming gain for each RF chain, it requires complex circuitry and consumes relatively high power. An alternative way is to connect each RF chain only with part of the antennas, i.e., the partially-connected architecture, which can greatly reduce the hardware cost and power consumption. However, the traditional HBF design algorithms for the fully-connected architecture cannot be straightforwardly applied to the partially-connected one as the analog beamforming matrix becomes a block diagonal matrix.  

Compared to its fully-connected counterpart, HBF for the partially-connected architecture has been less well studied. The first effort was in \cite{Yu:2016}, where the analog and digital precoders were alternatively optimized by updating one while fixing the other. A semi-definite relaxation based algorithm was proposed to optimize the digital precoder. However, the whole design was still based on the matrix factorization approach instead of directly minimizing the spectral efficiency. In \cite{Li:2017}, the authors first designed the analog precoder for high signal to noise ratio (SNR) and low SNR regions, respectively, and then applied the water-filling algorithm to optimize the digital precoder. However, only the narrowband scenario was considered. In \cite{Sohrabi:2017}, the authors considered the design of HBF for MIMO orthogonal frequency division multiplexing (OFDM) systems. By utilizing the average of the covariance matrices of frequency domain channels, the original algorithms proposed for the narrowband scenario can be extended to the broadband one. However, to solve the problem, the original objective function had to be replaced by an upper bound, which unavoidably led to performance loss.  

\subsection{Contributions and Paper Organization}\label{subsec:contribution}
In this paper, we investigate the HBF problem for mmWave massive MIMO-OFDM systems with the partially-connected architecture, aiming at maximizing the spectral efficiency. In contrast to the previous design approaches that adopt surrogate objectives, e.g., to consider a matrix factorization problem to approximate the fully digital beamformer \cite{Ayach:2014,Yu:2016} or to replace the original objective function by some bound or approximation \cite{Sohrabi:2016,Sohrabi:2017}, we directly tackle the spectral efficiency maximization (SEM) problem based on an equivalent matrix weighted sum mean square error minimization (WMMSE) problem. The main contributions of this paper are summarized as follows:

\begin{itemize}
\item
Inspired by previous works on the fully-digital beamforming design for narrowband systems \cite{Sampath:2001,Shi:2011,Christensen:2008}, we prove that the HBF beamforming for maximizing the spectral efficiency is equivalent to the WMMSE problem, which provides a new and promising algorithmic approach for the HBF optimization. We also show that this design approach is applicable to both the partially-connected and fully-connected architectures. 

\item
To deal with the difficulty in the highly non-convex and multivariate HBF optimization problem, we show that the WMMSE problem can be decomposed into the hybrid precoding and hybrid combining subproblems, where both the optimal digital precoder and combiner of the two subproblems have closed-form solutions. With the unit modulus constraint, the analog precoder and combiner are more challenging to optimize, for which we propose an element iteration (EI) algorithm and a MO based algorithms. Finally, an alternating optimization approach is applied which updates the hybrid precoder and combiner iteratively. The proposed WWMSE-EI and WMMSE-MO HBF optimization algorithms are proved to be able to make the spectral efficiency monotonously increase and thus converge. Simulation results show that the proposed HBF optimization algorithms can reduce the required SNR by around $2\text{dB}$ to achieve the same spectral efficiency when compared with the conventional algorithms.

\item
To reduce the computational complexity, we propose a low complexity MMSE-EI HBF optimization algorithm. We also show that the MMSE-EI algorithm can provide a good initialization point for the WMMSE-EI and WMMSE-MO algorithms to speed up their convergence and improve the spectral efficiency. Furthermore, we propose HBF optimization algorithms considering finite resolution phase shifters, which are shown via simulations to achieve higher spectral efficiently than the simple algorithm with uniform phase quantization. 


\end{itemize}

The remainder of this paper is organized as follows. In Section \ref{sec:sysMod-proFormu}, we introduce the mmWave MIMO-OFDM system model with the partially-connect architecture and formulate the HBF optimization problem. In Section \ref{sec:design}, we first prove the equivalence between the SEM HBF problem and the WMMSE HBF problem, and then propose two iterative algorithms for solving the WMMSE problem. In Section \ref{sec:modified-HBF-design}, we present some modified HBF algorithms by considering the computational complexity and finite resolution phase shifters. In Section \ref{sec:analysis}, we prove the convergence of the proposed iterative algorithms and analysis their computational complexity. Finally, we provide various simulation results in Section \ref{sec:simulation} and conclude the paper in Section~\ref{sec:conclusion}. 

\subsection{Notations}
Throughout this paper, $a$ (or $A$), ${\mathbf{a}}$ and ${\mathbf{A}}$ stand for a scaling factor, a column vector and a matrix, respectively. $\angle \left(  \cdot  \right)$ denotes the angle of a complex variable. ${\left(  \cdot  \right)^ * }$, ${\left(  \cdot  \right)^T}$ and ${\left(  \cdot  \right)^H}$ respectively represent the operation of the conjugate, transpose and conjugate transpose of ${\mathbf{a}}$ or ${\mathbf{A}}$. $\left|  \cdot  \right|$, ${\left\|  \cdot  \right\|_F}$, ${\text{Tr}}\left\{  \cdot  \right\}$ and ${\left(  \cdot  \right)^{ - 1}}$ denote the determinant (or module for a complex variable), Frobenius norm, trace and inverse of matrix ${\mathbf{A}}$. $\odot$ is the Hadamard product of two matrices.  ${\mathbf{A}}\left( {i,:} \right)$, ${\mathbf{A}}\left( {:,j} \right)$ and ${\mathbf{A}}\left( {i,j} \right)$ denote the $i$-th row, the $j$-th column and the element in the $i$-th row and the $j$-th column of ${\mathbf{A}}$, respectively. ${\mathbb{C}^{m \times n}}$ is a complex space with ${m \times n}$ dimensions and ${\mathbf{I}}_N$ denotes the $N\times N$ identity matrix. $\mathbb{E}[ \cdot ]$ denotes the expectation operation. Finally, ${\mathbf{x}} \sim \mathcal{C}\mathcal{N}( {{\boldsymbol{\mu}},{\mathbf{K}}} )$ means that ${\mathbf{x}}$ is a circularly symmetric complex Gaussian vector whose mean is ${\boldsymbol{\mu}}$ and covariance
matrix is ${\mathbf{K}}$. 

\section{System Model and Problem Formulation}\label{sec:sysMod-proFormu}
\begin{figure*}[!t]
\begin{center}
\centering
\includegraphics[width=0.99\textwidth]{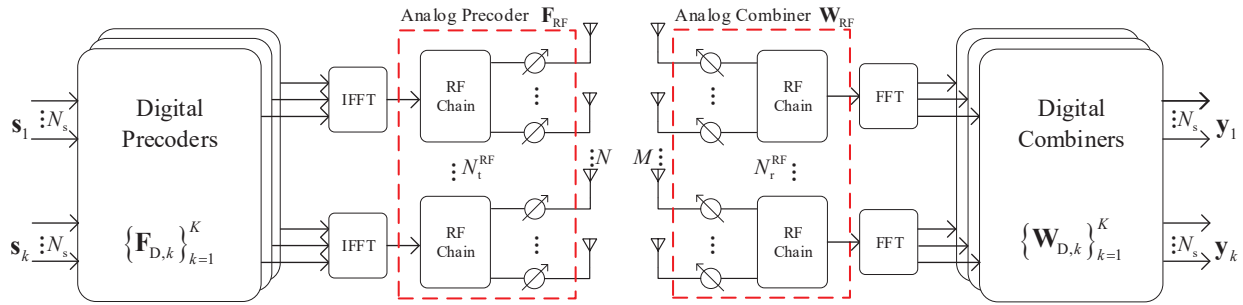}
\caption{Diagram of an mmWave MIMO-OFDM system with the partially-connected HBF architecture.}
\label{fig:system-model}
\end{center}
\end{figure*}
\subsection{System Model}\label{subsec:system-model}
Consider the downlink of an mmWave MIMO-OFDM system with the partially-connected HBF architecture, as shown in Fig. \ref{fig:system-model}. At the transmitter (i.e., the base station (BS)), a digital transmit beamformer (precoder) ${{\mathbf{F}}_{{\text{D,}}k}} \in {\mathbb{C}^{N_{\text{t}}^{{\text{RF}}} \times {N_{\text{s}}}}}$ is first employed to precode $N_\mathrm{s}$ data streams, denoted by vector ${{\mathbf{s}}_k} \in {\mathbb{C}^{{N_s} \times 1}}$ with $\mathbb{E}\left[ {{{\mathbf{s}}_k}{\mathbf{s}}_k^H} \right] = {{\mathbf{I}}_{{N_{\text{s}}}}}$, at the $k$-th subcarrier, for $k = 1,2, \ldots ,K$, where $N_{\mathrm{t}}^{\mathrm{RF}}$ denotes the number of transmit RF chains. Then, each of the $N_{\text{t}}^{\text{RF}}$ precoded streams are converted to the time domain by a $K$-point inverse fast Fourier transform (IFFT). After adding a cyclic prefix (omitted in Fig. \ref{fig:system-model} due to space limit), each stream is up-converted to the carrier frequency by passing through a dedicated RF chain. Before transmitting the RF signals at the $N$ antennas, an analog precoder consisting of a number of phase shifters is deployed for enhancing the beamforming gain. From an equivalent baseband point of view, the transmitted signal vector at the $k$-th subcarrier of the $N$ antennas is represented by ${{\mathbf{x}}_k} = {{\mathbf{F}}_{{\text{RF}}}}{{\mathbf{F}}_{{\text{D,}}k}}{{\mathbf{s}}_k}$, where ${{\mathbf{F}}_{{\text{RF}}}} \in {\mathbb{C}^{N \times N_{\text{t}}^{{\text{RF}}}}}$ denotes the analog precoder. Note that ${\mathbf{F}}_{{\text{RF}}}$ is the same for all the $K$ subcarriers because of its post-IFFT processing. Considering the maximum transmit power constraint per subcarrier, we have $\left\| {{{\mathbf{F}}_{{\text{RF}}}}{{\mathbf{F}}_{{\text{D,}}k}}} \right\|_F^2 \leqslant 1$.

At the receiver (i.e., the user equipment (UE)) with $M$ receive antennas, the equivalent baseband received signal at the $k$-th subcarrier can be expressed as ${{{\mathbf{\tilde y}}}_k} ={{\mathbf{H}}_k}{{\mathbf{F}}_{{\text{RF}}}}{{\mathbf{F}}_{{\text{D,}}k}}{{\mathbf{s}}_k} + {{\mathbf{n}}_k}$, where ${{\mathbf{n}}_k} \sim \mathcal{C}\mathcal{N}( {{\mathbf{0}},\sigma^2{{\mathbf{I}}_M}} )$ denotes the additive white Gaussian noise at the $k$-th subcarrier, and ${{\mathbf{H}}_k} \in {\mathbb{C}^{M \times N}}$ denotes the geometric model based channel matrix at the $k$-th subcarrier \cite{Yu:2016,Sohrabi:2017,Lin:2019}, which is given by 
\begin{equation}\label{eqn:channel-model}
{{\mathbf{H}}_k} = \sqrt {\frac{{MN}}{{{N_{\text{C}}}{N_{\text{R}}}}}} \sum\limits_{c = 1}^{{N_{\text{C}}}} {\sum\limits_{l = 1}^{{N_{\text{R}}}} {{h_{cl}}} } {{\mathbf{a}}_{\text{r}}}\left( {\theta _{cl}^{\text{r}}} \right){{\mathbf{a}}_{\text{t}}}{\left( {\theta _{cl}^{\text{t}}} \right)^H}{{\text{e}}^{ - {\text{j}}\frac{{2\pi }}{K}\left( {k - 1} \right)}},
\end{equation}
where ${{N_{\text{C}}}}$ and ${{N_{\text{R}}}}$ denote the number of clusters and the number of rays in each cluster, respectively. Likewise, ${{{h_{cl}}}}$, ${\theta_{cl}^{\text{r}}}$ and ${\theta_{cl}^{\text{t}}}$ represent the complex gain, the angles of arrival and departure (AoA and AoD) of the $l$-th ray in the $c$-th propagation cluster, respectively. In addition, ${{\mathbf{a}}_{\text{r}}}\left(  \cdot  \right)$ and ${{\mathbf{a}}_{\text{t}}}\left(  \cdot  \right)$ denote the array response vectors of the receiver and transmitter, respectively. For example, as for a half-wavelength spaced uniform liner array (ULA) with $N$ antennas at the transceiver, the array response vector can be represented as ${\mathbf{a}}\left( \theta  \right) = \frac{1}{{\sqrt N }}{\left[ {1,{\text{e}^{\text{j}\pi \sin \theta }}, \ldots ,{\text{e}^{\text{j}\left( {N - 1} \right)\pi \sin \theta }}} \right]^{T}}$.    

The received signal is first processed by an analog receive beamformer (combiner) ${{\mathbf{W}}_{{\mathrm{RF}}}} \in {\mathbb{C}^{M \times N_{\mathrm{r}}^{{\text{RF}}}}}$ and then down-converted to the baseband, where $N_{\mathrm{r}}^{{\text{RF}}}$ denotes the number of receive RF chains. Note that ${{\mathbf{W}}_{{\text{RF}}}}$ is also the same for all subcarriers as similar to ${{\mathbf{F}}_{{\text{RF}}}}$. After cyclic prefix  removal and fast Fourier transform (FFT), the $N_{\mathrm{r}}^{{\text{RF}}}$ baseband signal streams are then passing through a low-dimensional digital combiner ${{\mathbf{W}}_{{\text{D,}}k}} \in {\mathbb{C}^{N_{\text{r}}^{{\text{RF}}} \times {N_{\text{s}}}}}$ at subcarrier $k$ with the output given by
\begin{equation}
{{\mathbf{y}}_k} = {\mathbf{W}}_{k}^H{{\mathbf{H}}_k}{{\mathbf{F}}_{k}}{{\mathbf{s}}_k} + {\mathbf{W}}_{k}^H{{\mathbf{n}}_k},
\end{equation}
where ${{\mathbf{F}}_{k}} = {{\mathbf{F}}_{{\text{RF}}}}{{\mathbf{F}}_{{\text{D,}}k}}$ and ${{\mathbf{W}}_{k}} = {{\mathbf{W}}_{{\text{RF}}}}{{\mathbf{W}}_{{\text{D,}}k}}$. It is assumed that  $N_{\text{r}}^{{\text{RF}}} = {N_{\text{s}}}$ so that the digital combiner ${\mathbf{W}}_{{\text{D,}}k}$ is a square matrix. The scenario with  $N_{\text{r}}^{{\text{RF}}} > {N_{\text{s}}}$ will be discussed in Section~\ref{subsec:convergence}.

In this paper, we mainly focus on the partially-connected architecture for the analog transmit and receive beamformers and assume that each transmit RF chain is only connected with ${N}/{{N_\text{t}^{{\text{RF}}}}}$ antennas at the transmitter and each receive RF chain to ${M}/{{N_\text{r}^{{\text{RF}}}}}$ antennas at the receiver, as shown in Fig. \ref{fig:system-model}. As such, both  ${{\mathbf{F}}_{{\text{RF}}}}$ and ${{\mathbf{W}}_{{\text{RF}}}}$ become block-diagonal matrices 
\begin{subequations}
\begin{align}
{{\mathbf{F}}_{{\text{RF}}}} &= {\text{blkdiag}}\left( {{{\mathbf{f}}_1},{{\mathbf{f}}_2}, \cdots ,{{\mathbf{f}}_{N_{\text{t}}^{{\text{RF}}}}}} \right),\label{eqn:F_RF_block_diag}\\{{\mathbf{W}}_{{\text{RF}}}} &= {\text{blkdiag}}\left( {{{\mathbf{w}}_1},{{\mathbf{w}}_2}, \cdots ,{{\mathbf{w}}_{N_{\text{r}}^{{\text{RF}}}}}} \right),\label{eqn:W_RF_block_diag}
\end{align}
\end{subequations}
where ${{\mathbf{f}}_q} \in {\mathbb{C}^{\frac{N}{{N_{\text{t}}^{{\text{RF}}}}} \times 1}}$ for $q = 1,2, \ldots ,N_{\text{t}}^{{\text{RF}}}$ and ${{\mathbf{w}}_n} \in {\mathbb{C}^{\frac{M}{{N_{\text{r}}^{{\text{RF}}}}} \times 1}}$ for $n = 1,2, \ldots ,N_{\text{r}}^{{\text{RF}}}$. Since these two analog beamformers are implemented using phase shifters, the non-zero elements in ${{\mathbf{F}}_{{\text{RF}}}}$ and ${{\mathbf{W}}_{{\text{RF}}}}$ are subject to the constant modulus constraint. 

\subsection{Problem Formulation}\label{subsec:pro-formu}
The achievable spectral efficiency of the aforementioned system at the $k$-th subcarrier is given by
\begin{equation}\label{eqn:rate}
    {R_k} = \log \left| {{{\mathbf{I}}_{{N_{\text{s}}}}} + \sigma^{ - 2}{\mathbf{W}}_{k}^H{{\mathbf{H}}_k}{{\mathbf{F}}_{k}}{\mathbf{F}}_{k}^H{\mathbf{H}}_k^H{{\mathbf{W}}_{k}}{{\left( {{\mathbf{W}}_{k}^H{{\mathbf{W}}_{k}}} \right)}^{-1}}} \right|.
\end{equation}
In this work, we aim at maximizing the average spectral efficiency over the $K$ subcarriers subject to the transmit power constraint and the constant modulus constraint of the analog beamformers. The problem can be formulated as follows
\begin{equation}\label{prb:rate-broadband}
\begin{array}{cl}
\displaystyle{\maximize_{{{\mathbf{F}}_{{\text{RF}}}},{{\mathbf{F}}_{{\text{D,}}k}},{{\mathbf{W}}_{{\text{RF}}}},{{\mathbf{W}}_{{\text{D,}}k}}}} & \frac{1}{K}\sum\limits_{k = 1}^K {{R_k}}  \\
{\text{s}}{\text{.t}}{\text{.}} & \left\| {{{\mathbf{F}}_{{\text{D,}}k}}} \right\|_F^2 \leqslant \frac{{N_{\text{t}}^{{\text{RF}}}}}{N}, \quad \forall{k}\\ \quad & |\mathbf{f}_q(p)| = 1, \quad \forall{p}, \forall{q}\\\quad& 
|\mathbf{w}_n(m)| = 1, \quad \forall{m}, \forall{n},
\end{array}
\end{equation}
where $\mathbf{f}_q(p)$ and $\mathbf{w}_n(m)$ denote the $p$-th element in $\mathbf{f}_q$ and the $m$-th element in $\mathbf{w}_n$, respectively. The transmit power constraint in (\ref{prb:rate-broadband}) comes from the requirement of $\left\| {{{\mathbf{F}}_{{\text{RF}}}}{{\mathbf{F}}_{{\text{D,}}k}}} \right\|_F^2 \leqslant 1$ and the fact that ${\mathbf{F}}_{{\text{RF}}}^H{{\mathbf{F}}_{{\text{RF}}}} = \frac{N}{{N_{\text{t}}^{{\text{RF}}}}}{{\mathbf{I}}_{N_{\text{t}}^{{\text{RF}}}}}$ when ${{\mathbf{F}}_{{\text{RF}}}}$ is a block diagonal matrix according to (5a). Throughout this paper, we focus on the HBF optimization and assume that perfect channel state information is available.

\section{HBF Optimization with the Partially-connected Architecture}\label{sec:design}

For the highly non-convex and multivariate optimizaiton problem in (\ref{prb:rate-broadband}), it is very difficult to get the optimal solution. Our main idea is that instead of directly solving the problem, we formulate a WMMSE problem and show that it is equivalent to the original SEM problem. It is worth noting that although the WMMSE design approach has been considered in \cite{Sampath:2001, Shi:2011, Christensen:2008} for the fully-digital beamforming optimization and in \cite{Nguyen:2017,Lin:2019} for the HBF optimization, the equivalence in the HBF scenario has not been proved. In this section, we first formulate the WMMSE HBF optimization problem and prove its equivalence to the SEM problem. Then, to solve the WMMSE problem, we separate it into two subproblems, namely the hybrid precoding and hybrid combining optimization subproblems, and propose several effective algorithms. Finally, an alternating minimization algorithm is applied between the two subproblems for better performance. 


\subsection{The WMMSE Problem}\label{subsec:basic-idea}
Similar to that in \cite{Lin:2019}, we take the modified MSE as the performance metric and define the modified MSE matrix \cite{Joham:2015} as follows
\begin{equation}\label{eqn:modified-MSE-matrix}
\begin{aligned}
{\mathbf{E}_k} &= \mathbb{E}\left[ {\left( {{\mathbf{s}_k} - {\xi_k ^{ - 1}}{\mathbf{y}_k}} \right){{\left( {{\mathbf{s}_k} - {\xi _k^{ - 1}}{\mathbf{y}_k}} \right)}^H}} \right]\\
&= {{\mathbf{I}}_{{N_{\text{s}}}}} - \xi _k^{ - 1}{\mathbf{F}}_{k}^H{\mathbf{H}}_k^H{{\mathbf{W}}_{k}} - \xi _k^{ - 1}{\mathbf{W}}_{k}^H{{\mathbf{H}}_k}{{\mathbf{F}}_{k}} + \xi _k^{ - 2}\sigma^2{\mathbf{W}}_{k}^H{{\mathbf{W}}_{k}} + \xi _k^{ - 2}{\mathbf{W}}_{k}^H{{\mathbf{H}}_k}{{\mathbf{F}}_{k}}{\mathbf{F}}_{k}^H{\mathbf{H}}_k^H{{\mathbf{W}}_{k}},
\end{aligned}
\end{equation}
for $k=0,1,\ldots, K$, where $\xi_k$ a scaling factor to be jointly optimized with the hybrid beamformers~\cite{Lin:2019}. Defining a semi-positive definite matrix ${\boldsymbol{\Lambda}}_k\succeq {\bf{0}}$ as the weight matrix for the $k$-th subcarrier, the WMMSE problem can be formulated as 
\begin{equation}\label{prb:wmmse-broadband}
\begin{array}{cl}
\displaystyle{\minimize_{{{\bf{F}}_{{\text{RF}}}}, {\bf{F}}_{\text{D},k}, {{\bf{W}}_{{\text{RF}}}}, {{\bf{W}}_{{\text{D}},k}}, {\xi_k}, {\boldsymbol{\Lambda}}_k}} & \frac{1}{K}\sum\limits_{k = 1}^K {\left( {{\text{tr}}\left( {{{\boldsymbol{\Lambda}}_k} {{{\bf{ E}}}_k}} \right) - \log \left| {\boldsymbol{\Lambda}}_k \right|} \right)}  \\
{\text{s}}{\text{.t}}{\text{.}} &\left\| {{{\bf{F}}_{{\text{D,}}k}}} \right\|_F^2 \leqslant \frac{{N_{\text{t}}^{{\text{RF}}}}}{N}, \quad \forall{k}\\  \quad & |\mathbf{f}_q(p)| = 1, \quad \forall{p}, \forall{q}\\\quad& 
|\mathbf{w}_n(m)| = 1, \quad \forall{m}, \forall{n}.
\end{array}
\end{equation}
Fixing ${{\bf{F}}_{{\text{RF}}}}$,  ${{{\bf{F}}_{{\text{D,}}k}}}$, ${{\bf{W}}_{{\text{RF}}}}$, ${\boldsymbol{\Lambda}}_k$ and ${\xi_k}$, the solution of the optimal ${{\bf{W}}_{{\text{D}},k}}$ is given by
\begin{equation}\label{eqn:WD-opt}
    {\bf{W}}_{{\text{D}},k}^{{\text{mmse}}} = {\left( {{\bf{W}}_{{\text{RF}}}^H{{\bf{G}}_{k}}{\bf{G}}_{k}^H{{\bf{W}}_{{\text{RF}}}} + {\alpha _k}{{\bf{I}}_{N_{\text{r}}^{{\text{RF}}}}}} \right)^{ - 1}}{\bf{W}}_{{\text{RF}}}^H{{\bf{G}}_{k}},
\end{equation}
where $\mathbf{G}_{k}=\xi_{k}^{-1} \mathbf{H}_{k} \mathbf{F}_{k}$ and ${\alpha_{k}} = \frac{{\sigma^2\xi_k^{-2}M}} {{N_{\text{r}}^{{\text{RF}}}}}$. By substituting  ${\bf{W}}^{\text{mmse}}_{{\text{D}},k}$ back into (\ref{eqn:modified-MSE-matrix}), the corresponding MSE matrix becomes
\begin{equation}\label{eqn:MSE-matrix-with-optWD}
{\mathbf{E}}_k^{{\text{mmse}}} = {\left( {{{\mathbf{I}}_{{N_{\text{s}}}}} + \alpha_k^{ - 1}{\mathbf{G}}_k^H{{\mathbf{W}}_{{\text{RF}}}}{\mathbf{W}}_{{\text{RF}}}^H{{\mathbf{G}}_k}} \right)^{ - 1}}.
\end{equation}
Then, the following theorem will set up the equivalence between the WMMSE problem and the SEM problem in (\ref{prb:rate-broadband}). 

\begin{theorem}\label{theorem:rate-wmmse-eq}
The problem in (\ref{prb:wmmse-broadband}) is equivalent to the problem in (\ref{prb:rate-broadband}) in the sense that the global optimal solution of ${{\bf{F}}_{{\mathrm{RF}}}}$, ${{\bf{F}}_{{\mathrm{D,}}k}}$, and ${{\bf{W}}_{{\mathrm{RF}}}}$ for the two problems are identical.
\end{theorem}

\textit{Proof}: First the optimal ${\bf{W}}{}_{{\text{D}},k}$ can be acquired by differentiating the objective function of (\ref{prb:wmmse-broadband}) with ${\bf{W}}{}_{{\text{D}},k}$ and setting the result to zero, which is exactly the same as ${\bf{W}}_{{\text{D}},k}^{{\text{mmse}}}$ in (\ref{eqn:WD-opt}). As for the weight matrix ${\boldsymbol{\Lambda}}_k$, it can be shown in the same way that the optimal one has a closed-form expression
 \begin{equation}\label{Weighted-matrix}
     {\boldsymbol{\Lambda}}_k^{{\text{opt}}} = {\mathbf{E}}_k^{-1}.
 \end{equation}
With the optimal ${\bf{W}}{}_{{\text{D}},k}$ and ${\boldsymbol{\Lambda}}_k$, the problem in (\ref{prb:wmmse-broadband}) can be stated as
\begin{equation}\label{prb:wmmse-equ-broadband}
\begin{array}{cl}
\displaystyle{\minimize_{{{\bf{F}}_{{\text{RF}}}},{{\bf{F}}_{{\text{D,}}k}}{{\bf{W}}_{{\text{RF}}}}}} & { - \frac{1}{K}\sum\limits_{k = 1}^K {\log \left| {{{\left( {{\bf{E}}_k^{{\text{mmse}}}} \right)}^{ - 1}}} \right|} } \\
{\text{s}}{\text{.t}}{\text{.}} &\left\| {{{\bf{F}}_{{\text{D,}}k}}} \right\|_F^2 \leqslant \frac{{N_{\text{t}}^{{\text{RF}}}}}{N}, \quad \forall{k}\\  \quad & |\mathbf{f}_p(q)| = 1, \quad \forall{p}, \forall{q}\\\quad& 
|\mathbf{w}_m(n)| = 1, \quad \forall{m}, \forall{n},
\end{array}
\end{equation}
with 
\begin{equation}\label{eqn:WD-rate-MMSE}
 \begin{aligned}
   {\log \left| {{{\left( {{\bf{E}}_k^{{\text{mmse}}}} \right)}^{ - 1}}} \right|} & = \log \left| {{{\mathbf{I}}_{{N_{\text{s}}}}} + \alpha_k^{ - 1}{\mathbf{G}}_k^H{{\mathbf{W}}_{{\text{RF}}}}{\mathbf{W}}_{{\text{RF}}}^H{{\mathbf{G}}_k}} \right|\\ 
  &\mathop { = }\limits^{\left( {\text{a}} \right)} \log \left| {{{\mathbf{I}}_{{N_{\text{s}}}}} + \frac{{N_{\text{r}}^{{\text{RF}}}}}{{{\sigma ^2}M}}{\mathbf{W}}_{{\text{RF}}}^H{{\mathbf{H}}_k}{{\mathbf{F}}_k}{\mathbf{F}}_k^H{\mathbf{H}}_k^H{{\mathbf{W}}_{{\text{RF}}}}} \right|,
 \end{aligned}
\end{equation}
where ${\left( {\text{a}} \right)}$ follows from the definition of $\bf{G}_k$ and the Woodbury matrix identity that $\det \left( {{\bf{I}} + {\bf{XY}}} \right) = \det \left( {{\bf{I}} + {\bf{YX}}} \right)$. It can be seen that (\ref{eqn:WD-rate-MMSE}) is exactly the same as (\ref{eqn:rate}) for the case of $N_{\text{r}}^{{\text{RF}}} = {N_{\text{s}}}$. This shows that the problem of (\ref{prb:wmmse-broadband}) is equivalent to the problem (\ref{prb:rate-broadband}) in the sense that the global optimal solution of ${{\bf{F}}_{{\text{RF}}}}$, ${{\bf{F}}_{{\text{D,}}k}}$, and ${{\bf{W}}_{{\text{RF}}}}$ for the two
problems are identical. The proof is thus completed.  \hfill $\blacksquare$




Theorem \ref{theorem:rate-wmmse-eq} implies that the SEM problem can be achieved by solving the problem based on the WMMSE criterion. It is also worth noting that the proof of Theorem \ref{theorem:rate-wmmse-eq} can be shown to be applicable to both the fully-connected HBF architecture and the partially-connected one as there is no constraint to the structure of the analog beamformers in the proof. 

In the following subsections, we focus on solving the problem in (\ref{prb:wmmse-broadband}). Although it is almost intractable to straightly obtain the optimal solution of (\ref{prb:wmmse-broadband}) due to the coupled multiple variables and the non-convex constraints, we propose the following solution. First, we show that the original problem can be separated into the hybrid precoding and hybrid combining subproblems. Then, we show that both the optimal digital precoder and combiner have closed-form expressions when dealing with the two subproblems, and the analog precoder and combiner can be optimized with some iterative algorithms in a unified way. Finally, the alternative optimization between these two subproblems can be applied to iteratively update the hybrid beamformers for better performance.


\subsection{Hybrid Precoding Design}\label{subsec:precoding}
By first fixing the hybrid combiner ${{\bf{W}}_{{\text{RF}}}},{{\bf{W}}_{{\text{D}},k}}$ and the weight matrix ${\boldsymbol{\Lambda}}_k$ in (\ref{prb:wmmse-broadband}), we obtain the hybrid precoding optimization subproblem. By further fixing the analog precoder ${{\bf{F}}_{{\text{RF}}}}$ in this subproblem and acccording to the Karush-Kuhn-Tucker (KKT) conditions, the optimal ${{\mathbf{F}}_{{\text{D,}}k}}$ and $\xi_k$ has a closed-form solution as follows \cite{Lin:2019,Joham:2015}
\begin{equation}\label{eqn:optimal-FDk}
{\mathbf{F}}_{{\text{D}},k}^{{\text{opt}}} = \xi _k\widetilde{\mathbf{F}}_k^{ - 1}{\mathbf{F}}_{{\text{RF}}}^H{\widetilde{\mathbf{G}}_k^H}{\boldsymbol{\Lambda}_k}
\end{equation}
and 
\begin{equation}\label{eqn:Xi-optimal}
     {\xi _k} = 1/\sqrt{ {\frac{N}{{N_{\text{t}}^{{\text{RF}}}}}\|{\widetilde{\mathbf{F}}_k^{ - 1}{\mathbf{F}}_{{\text{RF}}}^H{\widetilde{\mathbf{G}}_k^H}{{\boldsymbol{\Lambda }}_k}}\|_F^2}},
\end{equation}
where we define 
\begin{equation}\label{eqn:Fd-parameters}
{\widetilde{\mathbf{G}}_k^H} = {\mathbf{H}}_k^H{{\mathbf{W}}_k}, 
\quad \beta_k= \frac{{\sigma^2NM{\text{tr}}\left( {{{\boldsymbol{\Lambda}}_k}{\mathbf{W}}_{{\text{D,}}k}^H{{\mathbf{W}}_{{\text{D,}}k}}} \right)}}{{N_{\text{t}}^{{\text{RF}}}N_{\text{r}}^{{\text{RF}}}}}, \quad \widetilde{\mathbf{F}}_k={{\mathbf{F}}_{{\text{RF}}}^H{\widetilde{\mathbf{G}}_k^H}{{\boldsymbol{\Lambda }}_k}\widetilde{\mathbf{G}}_k{{\mathbf{F}}_{{\text{RF}}}} + {\beta _k}{{\mathbf{I}}_{N_{\text{t}}^{{\text{RF}}}}}}. 
\end{equation}

By substituting (\ref{eqn:optimal-FDk}) and (\ref{eqn:Xi-optimal}) back into the objection function in (\ref{prb:wmmse-broadband}), and assuming a fixed hybrid combiner, the original problem is now reduced to the one for optimizing the analog precoder ${{\mathbf{F}}_{{\text{RF}}}}$ as follows
\begin{equation}\label{prb:analog-precoder}
\begin{array}{cl}
\displaystyle{\minimize_{\mathbf{F}_{\mathrm{RF}}}} & f\left( {{{\mathbf{F}}_{{\text{RF}}}}} \right) \\
{\text{s}}{\text{.t}}{\text{.}}&|\mathbf{f}_q(p)| = 1, \quad \forall{p}, \forall{q},
\end{array}
\end{equation}
where 
\begin{equation}
    f\left( {{{\mathbf{F}}_{{\text{RF}}}}} \right) = \frac{1}{K}\sum\limits_{k = 1}^K {{\text{tr}}\left( {{{\left( {{{\boldsymbol{\Lambda }}_k}^{ - 1} + \beta_k^{ - 1}{\widetilde{\mathbf{G}}_{k}}{{\mathbf{F}}_{{\text{RF}}}}{\mathbf{F}}_{{\text{RF}}}^H{{\widetilde{\mathbf{G}}_{k}^H}}} \right)}^{ - 1}}} \right)}.
\end{equation} 
However, the above problem is still non-convex and it is difficult to obtain the optimal solution. Thus, we propose two iterative algorithms for obtaining a local optimal solution. 

\subsubsection{The EI Algorithm}\label{subsubsec:EI-Algorithm}
We first propose an element-by-element iterative optimization algorithm. In each iteration, the contribution of an analog precoding element, say ${{\mathbf{f}}_q}\left( p \right) = {\text{e}^{\text{j}{\theta _{pq}}}}$ without loss of generality, to the objective function is derived and optimized when other elements are fixed. In particular, By defining ${\Omega _{q,k}} \triangleq {\boldsymbol{\Lambda }}_k^{ - 1} + \beta_k^{ - 1}{\widetilde{\mathbf{G}}_{k}}{\mathbf{\bar F}}_{{\text{RF}}}^q{\left( {{\mathbf{\bar F}}_{{\text{RF}}}^q} \right)^H}{{\widetilde{\mathbf{G}}_{k}^H}}$, ${{\mathbf{A}}_{q,k}} \triangleq \beta_k^{ - 1}{\widetilde{\mathbf{G}}_{k}^H}\Omega _{q,k}^{ - 2}{\widetilde{\mathbf{G}}_{k}}$ and ${{\mathbf{B}}_{q,k}} \triangleq \frac{{N_{\text{t}}^{{\text{RF}}}}}{N}{{\mathbf{I}}_N} + \beta_k^{ - 1}{{\widetilde{\mathbf{G}}_{k}^H}}\Omega _{q,k}^{ - 1}{\widetilde{\mathbf{G}}_{k}}$, where ${\mathbf{f}}_{{\text{RF}}}^q$ and ${\mathbf{\bar F}}_{{\text{RF}}}^q$ respectively denote ${{\mathbf{F}}_{{\text{RF}}}}\left( {:,q} \right)$ and a sub-matrix of ${{\mathbf{F}}_{{\text{RF}}}}$ after removing ${{\mathbf{F}}_{{\text{RF}}}}\left( {:,q}\right)$, the objective function in (\ref{prb:analog-precoder}) can be rewritten as 
\begin{equation}\label{fFRF}
\begin{aligned}
{f\left( {{{\mathbf{F}}_{{\text{RF}}}}} \right)}
& ={\frac{1}{K}\sum\limits_{k = 1}^K {{\text{tr}}\left( {{{\left( {{\Omega _{q,k}} + \beta_k^{ - 1}{\widetilde{\mathbf{G}}_{k}}{\mathbf{f}}_{{\text{RF}}}^q{{\left( {{\mathbf{f}}_{{\text{RF}}}^q} \right)}^H}{{\widetilde{\mathbf{G}}_{k}^H}}} \right)}^{ - 1}}} \right)} } \\{}&\mathop  = \limits^{(a)}\frac{1}{K}\sum\limits_{k = 1}^K {\left( {{\text{tr(}}\Omega _{q,k}^{ - 1}{\text{)}} - \frac{{{\text{tr(}}\beta_k^{ - 1}\Omega _{q,k}^{ - 1}{\widetilde{\mathbf{G}}_{k}}{\bf{f}}_{{\text{RF}}}^q{{\left( {{\bf{f}}_{{\text{RF}}}^q} \right)}^H}{\widetilde{\mathbf{G}}_{k}^H}\Omega _{q,k}^{ - 1}{\text{)}}}}{{1 + {\text{tr(}}\beta_k^{ - 1}\Omega _{q.k}^{ - 1}\widetilde{\mathbf{G}}_{k}{\bf{f}}_{{\text{RF}}}^q{{\left( {{\bf{f}}_{{\text{RF}}}^q} \right)}^H}{\widetilde{\mathbf{G}}_{k}^H}{\text{)}}}}} \right)}  \\{}&{\mathop  = \limits^{\left( b \right)} }\left( {\frac{1}{K}\sum\limits_{k = 1}^K {{\text{tr(}}\Omega _{q,k}^{ - 1}{\text{)}}} } \right) - \left( {\frac{1}{K}\sum\limits_{k = 1}^K {\frac{{{{\left( {{\bf{f}}_{{\text{RF}}}^q} \right)}^H}{{\bf{A}}_{q,k}}{\bf{f}}_{{\text{RF}}}^q}}{{{{\left( {{\bf{f}}_{{\text{RF}}}^q} \right)}^H}{{\bf{B}}_{q,k}}{\bf{f}}_{{\text{RF}}}^q}}} } \right),
\end{aligned}
\end{equation}
where (a) follows from the fact that ${{\left( \mathbf{X}+\mathbf{Y} \right)}^{-1}}={{\mathbf{X}}^{-1}}-\frac{{{\mathbf{X}}^{-1}}\mathbf{Y}{{\mathbf{X}}^{-1}}}{\text{1+tr}\left( {{\mathbf{X}}^{-1}}\mathbf{Y} \right)}$ for a full-rank matrix ${\mathbf{X}}$ and a rank-one matrix ${\mathbf{Y}}$, and (b) follows from the property of ${\text{tr}}\left( {{\mathbf{XY}}} \right) = {\text{tr}}\left( {{\mathbf{YX}}} \right)$. Note that $\text{e}^{\text{j}{\theta _{pq}}}$ is contained only in the last term of (\ref{fFRF}), which can be written in the following form
\begin{equation}\label{eqn:EI-term-theta}
-{\frac{1}{K}\sum\limits_{k = 1}^K {\frac{{{{\left( {{\bf{f}}_{{\text{RF}}}^q} \right)}^H}{{\bf{A}}_{q,k}}{\bf{f}}_{{\text{RF}}}^q}}{{{{\left( {{\bf{f}}_{{\text{RF}}}^q} \right)}^H}{{\bf{B}}_{q,k}}{\bf{f}}_{{\text{RF}}}^q}}} } = -\frac{1}{K}\sum\limits_{k = 1}^K {\frac{{{A_k} + {B_k}\cos \left( {{\theta _{pq}} + {\theta _{1,k}}} \right)}}{{{C_k} + {D_k}\cos \left( {{\theta _{pq}} + {\theta _{2,k}}} \right)}}},
\end{equation}
where 
\begin{equation}\label{eqn:EI-ABCD}
\begin{array}{*{20}{l}}
  {A_k} = {\mathbf{\hat f}}_{q,p}^H{{\mathbf{A}}_{q,k}}{{{\mathbf{\hat f}}}_{q,p}} + {{\mathbf{A}}_{q,k}}\left( {p,p} \right),\quad{B_k} = 2\left| {{\mathbf{\hat f}}_{q,p}^H{{\mathbf{A}}_{q,k}}\left( {:,p} \right)} \right|, \quad{\theta _{1,k}} = \angle \left( {{\mathbf{\hat f}}_{q,p}^H{{\mathbf{A}}_{q,k}}(:,p)} \right), \\ 
  {C_k} = {\mathbf{\hat f}}_{q,p}^H{{\mathbf{B}}_{q,k}}{{{\mathbf{\hat f}}}_{q,p}} + {{\mathbf{B}}_{q,k}}\left( {p,p} \right),\quad{D_k} = 2\left| {{\mathbf{\hat f}}_{q,p}^H{{\mathbf{B}}_{q,k}}\left( {:,p} \right)} \right|, \quad{\theta _{2,k}} = \angle \left( {{\mathbf{\hat f}}_{q,p}^H{{\mathbf{B}}_{q,k}}(:,p)} \right),
\end{array}
\end{equation}
where ${{{{\mathbf{\hat f}}}_{q,p}}}$ is defined as the resulting vector by only setting the element ${{\mathbf{f}}_q}\left( {p} \right)$ in the vector ${{\mathbf{f}}_{{\text{RF}}}^q}$ to zero without changing other elements. It can be found that (\ref{eqn:EI-term-theta}) is a one-dimensional scalar function of ${{\theta _{pq}}}$. However, due to the summation of the $K$ terms, the closed-form solution of the optimal ${{\theta _{pq}}}$ is difficult to derive. Fortunately, some simple but efficient search algorithms based on the golden section search or the simulated annealing algorithm \cite{Chenbaolin} can be applied to obtain the optimal ${{\theta _{pq}}}$ when fixing  other analog beamforming elements. 

The overall EI algorithm is summarized in Algorithm \ref{alg:1}. First, the contribution of $\theta_{pq}$ in ${{\mathbf{F}}_{{\text{RF}}}}$ to the objective function is derived in (\ref{eqn:EI-term-theta}). Then, $\theta _{pq}$ is optimized and updated via the one-dimensional search algorithm while guaranteeing that the objective function keeps decreasing. The element-by-element iterations will be repeated until a stop condition is satisfied finally. Therefore, the convergence of the analog precoder based on the EI algorithm can be strictly proved. 

\begin{algorithm}
\caption{The EI algorithm for the analog beamforming optimization with the partially-connected architecture}
\label{alg:1}
\hspace*{0.02in} {\bf Input:} ${\beta_k}$, $\boldsymbol{\Lambda }_k$, $ \sigma ^2$, ${{\widetilde{\mathbf{G}}_{k}}}$, $K$

\begin{algorithmic}[1]
\STATE {Initialize ${\mathbf{F}}_{{\text{RF}}}^{\left( i \right)}$ with random phases and set $i=0$};
\REPEAT
\FOR{$q = 1 \to N_{\text{t}}^{{\text{RF}}}$} 
\STATE Compute ${{\Omega }_{q,k}}$, ${{\mathbf{A}}_{q,k}}$ and ${{\mathbf{B}}_{q,k}}$ according to (\ref{fFRF});
\FOR{$p = \frac{{(q - 1)N}}{{N_{\text{t}}^{{\text{RF}}}}} + 1 \to \frac{{qN}}{{N_{\text{t}}^{{\text{RF}}}}}$}
\STATE Compute $A_k$, $B_k$, $C_k$, $D_k$, ${\theta _{1,k}}$ and ${\theta _{2,k}}$ according to (\ref{eqn:EI-ABCD});
\STATE Update ${\mathbf{F}}_{{\text{RF}}}^{\left( i \right)}(p,q)$ based on the one-dimensional search;
\ENDFOR
\ENDFOR
\STATE ${\mathbf{F}}_{{\text{RF}}}^{\left( {i + 1} \right)} = {\mathbf{F}}_{{\text{RF}}}^{\left( i \right)}$ and $i \leftarrow i + 1$;
\UNTIL a stop condition is satisfied.
\end{algorithmic}
\hspace*{0.02in} {\bf Output:} ${{\mathbf{F}}_{{\text{RF}}}}$
\end{algorithm}

\subsubsection{The MO Algorithm} \label{subsubsec:MO-algorithm}
The above EI algorithm involves alternating iterations among elements and thus may have high computational complexity. Here we propose the MO algorithm with less complexity. Note that the MO method has been applied to deal with a series of problems with the constant modulus constraint of the phase shifters \cite{Yu:2016,Lin:2019,Absil:2009}. The basic idea is to consider these optimization problems in a Riemannian manifold space, i.e., a complex circle manifold defined by the constant modulus constraint. Then, some gradient decent like algorithm can be adopted to iteratively update the optimization variable (the analog beamformer) in the direction of the Riemannian gradient. Finally, the updated optimization variable is retracted into a complex circle manifold to meet the the constant modulus constraint (For more detail, please refer to \cite{Yu:2016}). 

In the above design procedure, the crucial step is to derive the Riemannian gradient, which is the orthogonal projection of the Euclidean conjugate gradient. To the best of our knowledge, the MO method has not been applied to solve the HBF optimization problem with the partially-connected architecture. We show in this subsection that with the help of the following lemma, the Euclidean conjugate gradient in the partially-connected architecture can be derived.


\begin{lemma}\label{lem1}
For the partially-connected architecture, the Euclidean conjugate gradient $\nabla f\left( {{\mathbf{F}}_{\mathrm{RF}}} \right)$ of the real-valued function $f\left( {{\mathbf{F}}_{\mathrm{RF}}} \right)$ with respect to ${{{\mathbf{F}}_{{\mathrm{RF}}}}}$ can be given by
\begin{equation}\label{pca-gra}
\nabla f\left( {{{\mathbf{F}}_{{\mathrm{RF}}}}} \right) = {\nabla _{{\mathbf{F}}_{{\mathrm{RF}}}^*}}f\left( {{{\mathbf{F}}_{{\mathrm{RF}}}}} \right) \odot {{\mathbf{P}}_1},
\end{equation}
where ${\nabla _{{\mathbf{F}}_{{\mathrm{RF}}}^*}}f\left( {{{\mathbf{F}}_{{\mathrm{RF}}}}} \right) = \frac{{\partial f\left( {{{\mathbf{F}}_{{\mathrm{RF}}}}} \right)}}{{\partial {\mathbf{F}}_{{\mathrm{RF}}}^*}}$, ${{\mathbf{P}}_1} = {\mathrm{blkdiag}}( {{{\mathbf{p}}_1}, \ldots ,{{\mathbf{p}}_{N_{\mathrm{t}}^{{\mathrm{RF}}}}}} )$ is a block-diagonal matrix and ${{\mathbf{p}}_1} =  \ldots  = {{\mathbf{p}}_{N_{\mathrm{t}}^{{\mathrm{RF}}}}} = {{\mathbf{I}}_{\frac{N}{{N_{\mathrm{t}}^{{\mathrm{RF}}}}} \times 1}}$.
\end{lemma}

\textit{Proof}: See Appendix A. \hfill $\blacksquare$  
 
According to Lemma \ref{lem1}, the first step is to derive ${{\nabla }_{\mathbf{F}_{\text{RF}}^{*}}}f\left( {{\mathbf{F}}_{\text{RF}}} \right)$  without considering the partially-connected architecture. By applying the properties of the matrix differentiation \cite{Zhangxianda}, we have 
\begin{equation}\label{eqn:differentiation-f-RF}
\begin{aligned}
  {{\text{d}}\left( {f\left( {{{\mathbf{F}}_{{\text{RF}}}}} \right)} \right)}& ={{\text{d}}\left( {\frac{1}{K}\sum\limits_{k = 1}^K {{\text{tr}}\left( {{\mathbf{M}}_k^{ - 1}} \right)} } \right)} \\ 
  {}&{\mathop  = \limits^{(a)} }{ - {\text{tr}}\left( {\frac{1}{K}\sum\limits_{k = 1}^K {{\mathbf{M}}_k^{ - 2}{\text{d}}\left( {{{\mathbf{M}}_k}} \right)} } \right)} \\ 
  {}&{\mathop  = \limits^{(b)} }{ - {\text{tr}}\left( {\frac{1}{K}\sum\limits_{k = 1}^K {\beta_k^{ - 1}{{\widetilde{\mathbf{G}}_{k}}}{\mathbf{M}}_k^{ - 2}{\widetilde{\mathbf{G}}_{k}}^H{{\mathbf{F}}_{{\text{RF}}}}} {\text{d}}\left( {{\mathbf{F}}_{{\text{RF}}}^H} \right)} \right)}, 
\end{aligned}
\end{equation}
where (a) follows by defining ${{\mathbf{M}}_k} = {\boldsymbol{\Lambda }}_k^{ - 1} + \beta_k^{ - 1}{\widetilde{\mathbf{G}}_{k}}{{\mathbf{F}}_{{\text{RF}}}}{\mathbf{F}}_{{\text{RF}}}^H{{\widetilde{\mathbf{G}}_{k}^H}}$ and noting that ${\text{d}}\left( {{\text{tr}}\left( {{{\mathbf{X}}^{ - 1}}} \right)} \right) =  - {\text{tr}}\left( {{{\mathbf{X}}^{ - 2}}{\text{d}}\left( {\mathbf{X}} \right)} \right)$, and (b) follows from the basic properties of the matrix differentiation and the trace, i.e., $\text{d}\left( \mathbf{AXB} \right)=\mathbf{A}\text{d}\left( \mathbf{X} \right)\mathbf{B}$ ($\mathbf{A}$ and $\mathbf{B}$ are constant matrices independent of $\mathbf{X}$) and ${\text{tr}}\left( {{\mathbf{XY}}} \right) = {\text{tr}}\left( {{\mathbf{YX}}} \right)$. According to the relationship between the matrix differentiation of the scaling function and the Euclidean gradient \cite{Zhangxianda}, we have 
\begin{equation}\label{eqn:relation-diff-EucliGrad}
{\text{d}}\left( {f\left( {{{\mathbf{F}}_{{\text{RF}}}}} \right)} \right) = {\text{tr}}\left( {{\nabla _{{\mathbf{F}}_{{\text{RF}}}^*}}f\left( {{{\mathbf{F}}_{{\text{RF}}}}} \right){\text{d}}\left( {{\mathbf{F}}_{{\text{RF}}}^H} \right)} \right).  
\end{equation}
By comparing (\ref{eqn:differentiation-f-RF}) and (\ref{eqn:relation-diff-EucliGrad}), we obtain ${\nabla _{{\mathbf{F}}_{{\text{RF}}}^*}}f\left( {{{\mathbf{F}}_{{\text{RF}}}}} \right)$. According to Lemma \ref{lem1}, we finally have
\begin{equation}\label{eqn:CG}
\nabla f\left( {{{\mathbf{F}}_{{\text{RF}}}}} \right) =  - \left( {\frac{1}{K}\sum\limits_{k = 1}^K {\beta_k^{ - 1}{{\widetilde{\mathbf{G}}_{k}^H}}{\mathbf{M}}_k^{ - 2}\widetilde{\mathbf{G}}_{k}{{\mathbf{F}}_{{\text{RF}}}}} } \right) \odot {{\mathbf{P}}_1}.
\end{equation}
With the derived Euclidean conjugate gradient, the next step is to project it onto the tangent space to obtain the Riemannian gradient and update ${{{\mathbf{F}}_{{\text{RF}}}}}$ with a proper step size determined by the well-known Armijo backtracking algorithm. Finally, the retraction operation is applied to make the result satisfy the constant modulus constraint. The overall algorithm for the analog precoding with the partially-connected architecture is summarized in Algorithm \ref{alg:2}.
\begin{algorithm}[t]
\caption{The MO algorithm for the analog beamforming optimization with the partially-connected architecture}
\label{alg:2}
\hspace*{0.02in} {\bf Input:} ${\beta_k}$, $\boldsymbol{\Lambda }_k$, $ \sigma ^2$, ${{\widetilde{\mathbf{G}}_{k}}}$, $K$
\begin{algorithmic}[1]
\STATE Initialize ${\mathbf{F}}_{{\text{RF}}}^{\left( i \right)}$ with random phases and set $i=0$;
\REPEAT
\STATE Compute $\nabla f\left( {{\mathbf{F}}_{\text{RF}}} \right)$ according to Lemma \ref{lem1};
\STATE Update $\mathbf{F}_{\text{RF}}^{\left( i+1 \right)}$ based on the MO method;
\STATE $i\leftarrow i+1$;
\UNTIL a stopping condition is satisfied.
\end{algorithmic}\label{alg:manifold}
\hspace*{0.02in} {\bf Output:} ${{\mathbf{F}}_{{\text{RF}}}}$
\end{algorithm}

It is worth noting that the MO method can guarantee the convergence to a critical point where the gradient is zero according to Theorem 4.3.1 in \cite{Absil:2009}. Moreover, the well-developed conjugate gradient descent algorithm using the Armijo backtracking line search step and the Polak-Ribiere parameter can also ensure the objective function not to increase in each iteration \cite{Bertsekas:1999}. Thus, the  entire MO analog beamforming optimization algorithm converges, which further ensures that the iterations of the analog precoding optimization make the WMMSE not to increase until the stop condition is satisfied.

\subsection{Hybrid Combining Design}\label{subsec:hyd-combining}
In the above subsection, we have investigated the hybrid precoding optimization subproblem by assuming the hybrid combiner is fixed. Now we focus on the design of the hybrid combiner with a fixed precoder. Back to the original WMMSE problem in (\ref{prb:wmmse-broadband}), by fixing ${{\bf{F}}_{{\text{RF}}}}, {\bf{F}}_{\text{D},k},  {\xi_k}$ and ${\boldsymbol{\Lambda}}_k$, the objective function is only a function of ${{\bf{W}}_{{\text{RF}}}}$ and ${\bf{W}}_{\text{D},k}$. The closed-form solution of  the optimal ${{\bf{W}}_{{\text{D}},k}}$ has been given by (\ref{eqn:WD-opt}). By substituting it into the object function in (\ref{prb:wmmse-broadband}) and neglecting the second unrelated term, the optimization problem for  ${{\mathbf{W}}_{{\text{RF}}}}$ can be formulated as 
\begin{equation}\label{prb:analog-combiner}
\begin{array}{cl}
\displaystyle{\minimize_{\mathbf{W}_{\mathrm{RF}}}} & g\left( {{{\mathbf{W}}_{{\text{RF}}}}} \right) \\
{\text{s}}{\text{.t}}{\text{.}}&|{{\mathbf{w}}_n}(m)| = 1,\quad \forall m,\forall n, 
\end{array}
\end{equation}
where
\begin{equation}
g\left( {{{\mathbf{W}}_{{\text{RF}}}}} \right) = \frac{1}{K}\sum\limits_{k = 1}^K {{\text{tr}}\left( {{{\left( {{\boldsymbol{\Lambda}}_k^{ - 1} + \alpha_{k}^{ - 1}{\boldsymbol{\Lambda}}_k^{ - 1}{\mathbf{G}}_{k}^H{{\mathbf{W}}_{{\text{RF}}}}{\mathbf{W}}_{{\text{RF}}}^H{{\mathbf{G}}_{k}}} \right)}^{ - 1}}} \right)}.
\end{equation}
This problem is very similar to the analog precoding optimization problem in (\ref{prb:analog-precoder}) and thus can be solved in the same way as that in Section \ref{subsec:precoding}. In particular, by similarly defining  
${\Omega _{n,k}} \triangleq {\boldsymbol{\Lambda}}_k^{ - 1} + \alpha_{k}^{ - 1}{\boldsymbol{\Lambda}}_k^{ - 1}{\mathbf{G}}_{k}^H{\mathbf{\bar W}}_{{\text{RF}}}^n{\left( {{\mathbf{\bar W}}_{{\text{RF}}}^n} \right)^H}{{\mathbf{G}}_{k}}$, ${{\mathbf{A}}_{n,k}} \triangleq \alpha _{k}^{ - 1}{{\mathbf{G}}_{k}}\Omega _{n,k}^{ - 2}{\boldsymbol{\Lambda}}_k^{ - 1}{\mathbf{G}}_{k}^H$ and ${{\mathbf{B}}_{n,k}} \triangleq \frac{{N_{\text{r}}^{{\text{RF}}}}}{M}{{\mathbf{I}}_M}{\text{  +  }}\alpha _{k}^{ - 1}{{\mathbf{G}}_{k}}\Omega _{n,k}^{ - 1}{\boldsymbol{\Lambda}}_k^{ - 1}{\mathbf{G}}_{k}^H$, where ${\mathbf{w}}_{{\text{RF}}}^n$ and ${\mathbf{\bar W}}_{{\text{RF}}}^n$ respectively denote ${{\mathbf{W}}_{{\text{RF}}}}\left( {,:n} \right)$ and a sub-matrix of ${{\mathbf{W}}_{{\text{RF}}}}$ after removing ${{\mathbf{W}}_{{\text{RF}}}}\left( {,:n}\right)$, Algorithm \ref{alg:1} can be applied to optimize ${{\mathbf{W}}_{{\text{RF}}}}$ in the same way as that for ${{\mathbf{F}}_{{\text{RF}}}}$.

Similarly,  for the MO algorithm, the Euclidean gradient of the objective function of ${{{\mathbf{W}}_{{\text{RF}}}}}$ can be derived to be
\begin{equation}
\nabla g\left( {{{\mathbf{W}}_{{\text{RF}}}}} \right) =  - \left( {\frac{1}{K}\sum\limits_{k = 1}^K {\alpha_{k}^{ - 1}{{\mathbf{G}}_{k}}{\mathbf{N}}_k^{ - 2}{\mathbf{G}}_{k}^H{{\mathbf{W}}_{{\text{RF}}}}} } \right) \odot {{\mathbf{P}}_2},
\end{equation}
where ${{\mathbf{N}}_k} \triangleq {\boldsymbol{\Lambda}}_k^{ - 1} + \alpha _{k}^{ - 1}{\boldsymbol{\Lambda}}_k^{ - 1}{\mathbf{G}}_{k}^H{{\mathbf{W}}_{{\text{RF}}}}{\mathbf{W}}_{{\text{RF}}}^H{{\mathbf{G}}_{k}}$ and ${{\mathbf{P}}_2} = {\text{blkdiag}}\left( {{{{\mathbf{p'}}}_1}, \ldots ,{{{\mathbf{p'}}}_{N_{\text{r}}^{{\text{RF}}}}}} \right)$ with ${{\mathbf{p'}}_1} =  \ldots  = {{\mathbf{p'}}_{N_{\text{r}}^{{\text{RF}}}}} = {{\mathbf{I}}_{\frac{M}{{N_{\text{r}}^{{\text{RF}}}}} \times 1}}$. Thus, the Algorithm \ref{alg:2} can be applied to optimize the analog combiner. 

\subsection{Alternating Optimization for HBF}\label{subsec:alternating}
With the alternative optimization, the hybrid precoder and hybrid combiner can be jointly optimized by alternatively and iteratively using the proposed EI or MO algorithm in the above two subsections. Besides, another crucial step is to optimize the weight matrix ${{\boldsymbol{\Lambda}}_k}$. According to Theorem \ref{theorem:rate-wmmse-eq}, there exists a closed-form solution, i.e., ${\boldsymbol{\Lambda}}_k^{\text{opt}} = {\mathbf{E}}_k^{ - 1}$. Hence, we come up to the whole optimization process, which consists of three steps. 

Without loss of generality, we assume that each iteration starts with the optimization of the precoder. Thus, in the first step, by fixing the hybrid combiner and the weight matrix, ${{\mathbf{F}}_{{\text{RF}}}}$ and ${{\mathbf{F}}_{{\text{D}},k}}$ along with $\xi_k$ are optimized according to Algorithm 1 (or Algorithm 2), (\ref{eqn:optimal-FDk}) and (\ref{eqn:Xi-optimal}), respectively. Then, in the second step, by fixing the hybrid precoder and the weight matrix, ${{\mathbf{W}}_{{\text{RF}}}}$ and ${{\mathbf{W}}_{{\text{D}},k}}$ are optimized based Algorithm 1 (or Algorithm 2) and (\ref{eqn:WD-opt}), respectively. In the last step, ${{\boldsymbol{\Lambda}}_k}$, which is related to the SEM, is obtained according to (\ref{Weighted-matrix}). The three steps are repeated until the stop condition is satisfied. The overall HBF optimization for mmWave MIMO-OFDM systems with the partially-connected architecture is summarized in Algorithm \ref{alg:alternating}. We refer to the whole algorithm as the WMMSE-EI or WMMSE-MO algorithm according to whether the EI algorithm (i.e., Algorithm \ref{alg:1}) or MO algorithm (i.e., Algorithm \ref{alg:2}) is used when optimizing the analog beamformer. It is worth noting that if each iteration starts with the optimization of the combiner, the weight matrix should be updated before the optimization of the precoder to ensure the convergence of the spectral efficiency. The detailed proof of the convergence of the proposed WMMSE-EI and WMMSE-MO algorithms will be provided in Section \ref{subsec:convergence}. 

\begin{algorithm}[t]
\caption{HBF optimization with alternating minimization for mmWave MIMO-OFDM Systems with the partially-connected architecture}
\label{alg:alternating}
\hspace*{0.02in} {\bf Input:} ${{\mathbf{H}}_k}$ and ${\sigma ^2}$ 
\begin{algorithmic}[1]
\STATE Initialize ${\mathbf{W}}_{{\text{RF}}}^{\left( i \right)}$, ${\mathbf{F}}_{{\text{RF}}}^{\left( i \right)}$, ${\mathbf{W}}_{{\text{D}},k}^{\left( i \right)}$, ${\boldsymbol{\Lambda}}_k^{\left( i \right)}$ and $i = 0$.
\REPEAT
\STATE Compute ${\mathbf{F}}_{{\text{RF}}}^{\left( i \right)}$ based on Algorithm 1 (or Algorithm 2);
\STATE Compute ${\mathbf{F}}_{{\text{D}},k}^{\left( i \right)}$ according to ({\ref{eqn:optimal-FDk}});
\STATE Compute ${\mathbf{W}}_{{\text{RF}}}^{\left( i \right)}$ based on Algorithm 1 (or Algorithm 2);
\STATE Compute ${\mathbf{W}}_{{\text{D}},k}^{\left( i \right)}$ according to ({\ref{eqn:WD-rate-MMSE}});
\STATE Set ${\boldsymbol{\Lambda}}_k^{\left( i \right)} = {\mathbf{ E}}_k^{ - 1}$ based on (\ref{eqn:MSE-matrix-with-optWD});
\STATE $i \leftarrow i + 1$;
\UNTIL a stopping condition is satisfied;
\end{algorithmic}
\hspace*{0.02in} {\bf Output:} ${{\mathbf{F}}_{{\text{RF}}}}$,${{\mathbf{F}}_{{\text{D,}}k}}$,${{\mathbf{W}}_{{\text{RF}}}}$,${{\mathbf{W}}_{{\text{D,}}k}}$ 
\end{algorithm}

\section{Modified HBF Design Algorithms} \label{sec:modified-HBF-design}
In this section, we first consider the HBF optimization by modifying the WMMSE objective function to the MMSE and obtain a low complexity algorithm. We also modify the proposed WMMSE-EI algorithm with the practical consideration of finite resolution phase shifters.

\subsection{HBF Design Based on the MMSE Criterion}\label{subsec:MMSE}
In the previous section, we have investigated the HBF design for the partially-connected architecture aiming at maximizing the spectral efficiency by solving an equivalent WMMSE problem. We now consider its special case when the weight matrix is reduced to an identity matrix, i.e., without any weight. We refer to it as the MMSE HBF design. The motivation for the MMSE HBF design comes from three aspects. First, it can be regarded as a low-complexity version of the WMMSE design as the weight matrix does not need to be optimized. Second, it can be regarded as an initialization step for the WMMSE design by providing some good initial HBF matrices instead of random initialization for the WMMSE based algorithm. Third, in some cases, when a practical system is constrained to some particular modulation and coding scheme instead of the Gaussian code, the MSE metric becomes a direct performance measure to characterize the transmission reliability.


By setting $\boldsymbol{\Lambda}_{k}={\mathbf{I}}_{{N_{\text{s}}}}$ in (\ref{prb:wmmse-broadband}), the MMSE HBF optimization problem is formulated as 
\begin{equation}\label{prb:mmse-broadband}
\begin{array}{cl}
\displaystyle{\minimize_{{{\bf{F}}_{{\text{RF}}}}, {\bf{F}}_{\text{D},k}, {{\bf{W}}_{{\text{RF}}}}, {{\bf{W}}_{{\text{D}},k}}, {\xi_k}}} & \frac{1}{K}\sum\limits_{k = 1}^K {\left( {{\text{tr}}\left( { {{{\bf{ E}}}_k}} \right) } \right)}  \\
{\text{s}}{\text{.t}}{\text{.}} &\left\| {{{\bf{F}}_{{\text{D,}}k}}} \right\|_F^2 \leqslant \frac{{N_{\text{t}}^{{\text{RF}}}}}{N}, \quad \forall{k}\\  \quad & |\mathbf{f}_q(p)| = 1, \quad \forall{p}, \forall{q}\\\quad& 
|\mathbf{w}_n(m)| = 1, \quad \forall{m}, \forall{n}.
\end{array}
\end{equation}
Similar to the design approach for the WMMSE HBF problem, it can be separated into the hybrid precoding and combining subproblems. It can be shown that the digital precoder ${{\mathbf{F}}_{{\text{D}},k}}$, the scaling factor $\xi_k$, and the digital combiner ${{\mathbf{W}}_{{\text{D}},k}}$ can be expressed in the same form as those in (\ref{eqn:optimal-FDk}), (\ref{eqn:Xi-optimal}), and (\ref{eqn:WD-opt}) by just replacing $\boldsymbol{\Lambda}_{k}$ by ${\mathbf{I}}_{{N_{\text{s}}}}$ in them. Furthermore, the optimization problems for the analog precoder and combiner are similar and can be solved in the same way. We take the analog precoder optimization problem with the MMSE criterion for example, which is expressed as follows by replacing $\boldsymbol{\Lambda}_{k}$ by ${\mathbf{I}}_{{N_{\text{s}}}}$ in (\ref{prb:analog-precoder})


\begin{equation}\label{prb:analog-precoder-mmse}
\begin{array}{cl}
\displaystyle{\minimize_{\mathbf{F}_{\mathrm{RF}}}} &J\left( {{{\mathbf{F}}_{{\text{RF}}}}} \right)\\
{\text{s}}{\text{.t}}{\text{.}}&|{{\mathbf{f}}_q}(p)| = 1,\quad \forall p,\forall q, 
\end{array}
\end{equation}
with
\begin{equation}
J\left( {{{\mathbf{F}}_{{\text{RF}}}}} \right) = \frac{1}{K}\sum\limits_{k = 1}^K {\operatorname{tr} \left( {{{\left( {{{\mathbf{I}}_{{N_s}}} + \phi _k^{ - 1}{{\widetilde{\mathbf{G}}}_k}{{\mathbf{F}}_{{\text{RF}}}}{\mathbf{F}}_{{\text{RF}}}^H{\widetilde{\mathbf{G}}}_k^H} \right)}^{ - 1}}} \right)}, 
\end{equation}
where ${\phi _{k}} \triangleq \frac{{\sigma ^2NM}}{{N_{\text{t}}^{{\text{RF}}}N_{\text{r}}^{{\text{RF}}}}}{\text{tr}}\left( {{\mathbf{W}}_{{\text{D,}}k}^H{{\mathbf{W}}_{{\text{D,}}k}}} \right)$. Although it can be solved using the EI or MO algorithm in Section \ref{sec:design}, some new low complexity algorithm can be found. One way is to replace the objective function by one of its upper bounds. Using the Courant-Fisher min-max theorem \cite{Lin:2019,Horn:2012}, an upper bound of the objective function $J\left( {{{\mathbf{F}}_{{\text{RF}}}}} \right)$ in (\ref{prb:analog-precoder-mmse}) can be derived as follows
\begin{equation}\label{FRF-upperbound}
\begin{aligned}
    J\left( {{{\mathbf{F}}_{{\text{RF}}}}} \right) & \leqslant \frac{1}{K} \operatorname{tr} \left( {{\mathbf{F}}_{{\text{RF}}}^H{{\sum\limits_{k = 1}^K {\left( {\frac{{N_{\text{t}}^{{\text{RF}}}}}{N}{{\mathbf{I}}_{{N_{\text{t}}}}} + \phi _k^{ - 1}{\widetilde{\mathbf{G}}_{k}}^H{{\widetilde{\mathbf{G}}_{k}}}} \right)} }^{ - 1}}{{\mathbf{F}}_{{\text{RF}}}}} \right),
\end{aligned}
\end{equation}
which is defined as $J_{\text{UB}}\left( {{{\mathbf{F}}_{{\text{RF}}}}} \right)$. Using the matrix inversion equality and after some manipulation, we have 
\begin{equation}\label{equ:mmse-UB}
{J_{{\text{UB}}}}\left( {{{\mathbf{F}}_{{\text{RF}}}}} \right) = \frac{1}{K}\operatorname{tr} \left( {\left( {\sum\limits_{k = 1}^K {\frac{N}{{N_{\text{t}}^{{\text{RF}}}}}} {\mathbf{F}}_{{\text{RF}}}^H{{\mathbf{F}}_{{\text{RF}}}}} \right) - \frac{N}{{N_{\text{t}}^{{\text{RF}}}}}{\mathbf{F}}_{{\text{RF}}}^H{\mathbf{A}}{{\mathbf{F}}_{{\text{RF}}}}} \right) = \frac{{{N^2}}}{{N_{\text{t}}^{{\text{RF}}}}} - \frac{N}{{KN_{\text{t}}^{{\text{RF}}}}}{\text{tr}}\left( {{\mathbf{F}}_{{\text{RF}}}^H{\mathbf{A}}{{\mathbf{F}}_{{\text{RF}}}}} \right),
\end{equation}
where ${\mathbf{A}} = \sum\limits_{k = 1}^K {\left( {\phi _{k}^{ - 1}{{\widetilde{\mathbf{G}}_{k}}^H}{{\left( {\frac{{N_{\text{t}}^{{\text{RF}}}}}{N}{{\mathbf{I}}_{{N_{\text{s}}}}} + \phi _{k}^{ - 1}{\widetilde{\mathbf{G}}_{k}}{{\widetilde{\mathbf{G}}_{k}}}^H} \right)}^{ - 1}}{\widetilde{\mathbf{G}}_{k}}} \right)} $. The ${J}_{\text{UB}}$ minimization problem is then equivalent to the following problem
\begin{equation}\label{prb:MMSE-EI-max}
\begin{array}{cl}
\displaystyle{\maximize_{\mathbf{F}_{\mathrm{RF}}}} &{\operatorname{tr} \left( {{\mathbf{F}}_{{\text{RF}}}^H{\mathbf{A}}{{\mathbf{F}}_{{\text{RF}}}}} \right)}\\
{\text{s}}{\text{.t}}{\text{.}}&|{{\mathbf{f}}_q}(p)| = 1,\quad \forall p,\forall q. 
\end{array}
\end{equation}
Either the EI algorithm or the MO algorithm can be used to solve (\ref{prb:MMSE-EI-max}). From the low computational complexity point of view, we focus on the EI algorithm which can be shown to have a closed-form solution of $\mathbf{F}_\text{RF}$. Following the element-by-element optimization approach of the EI algorithm and taking the element of ${{\mathbf{f}}_q}\left( p \right) = {\text{e}^{\text{j}{\theta _{pq}}}}$ for example, the objective function in (\ref{prb:MMSE-EI-max}) can be expressed as the summation of  two terms with or without $\theta _{pq}$. That is, 
\begin{equation}\label{eqn:MMSE-objective-EI-max}
    {\operatorname{tr} \left( {{\mathbf{F}}_{{\text{RF}}}^H{\mathbf{A}}{{\mathbf{F}}_{{\text{RF}}}}} \right)} = 2B\cos \left( {{\theta _1} + {\theta _{pq}}} \right) + C,
\end{equation}
where $B = 2\left| {{\mathbf{\hat f}}_{q,p}^H{\mathbf{A}}\left( {:,p} \right)} \right|$, ${\theta _1} = \angle \left( {{\mathbf{\hat f}}_{q,p}^H{\mathbf{A}}\left( {:,p} \right)} \right)$, and $C$ is a term unrelatted to $\theta_{pq}$. It is not hard to find that $\theta _{pq}^{{\text{opt}}} =   - {\theta _1}$.

Similarly, it can be found that the above low complexity EI algorithm can be applied to the optimization of the analog combiner. Finally, by using the alternating minimization for the joint hybrid precoding and combining optimization until the stop conditional is satisfied, we obtain the optimized hybrid precoder and combiner. We refer to this HBF optimization algorithm as the MMSE-EI algorithm. Our computational complexity analysis in Section \ref{sec:analysis} along with the simulation results in Section  \ref{sec:simulation} will show that the MMSE-EI algorithm can reduce the computational complexity by  more than $90\%$ at the cost of less than $0.5\text{bits/s/Hz}$ in the spectral efficiency when compared to the WMMSE-EI and WMMSE-MO algorithms.

\subsection{HBF Design with Finite Resolution Phase Shifters}\label{subsec:phase-shifter}
In practical systems, as the phase shifters may have finite resolution, it is necessary to consider the HBF design in this scenario. According to \cite{Chige:2017}, a simple but efficiency way is to design the analog beamforming matrix with the assumption of infinite resolution, for example using the proposed WMMSE-EI, WMMSE-MO, and MMSE-EI algorithms, and then project the resulting phases into the quantized phase shifts set. We refer to this design approach as the projection based approach. 


However, for the EI related HBF algorithms, some specific optimization with finite resolution phase shifters can be conducted. As the analog precoding and combining optimization problem can be formulated and processed in a unified way, we take the analog precoding optimization as example. The original precoding optimization problem (\ref{prb:analog-precoder}) with infinite resolution can now be formulated as the following one with quantized phase shifts
\begin{equation}\label{prb:analog-precoder-Q}
\begin{array}{cl}
\displaystyle{\minimize_{\mathbf{F}_{\mathrm{RF}}}} & f\left( {{{\mathbf{F}}_{{\text{RF}}}}} \right) \\
{\text{s}}{\text{.t}}{\text{.}}&{{\mathbf{f}}_q}\left( {p} \right) \in \mathcal{F}, \quad \forall{p}, \forall{q},
\end{array}
\end{equation}
where $\mathcal{F} = \{ {{f^0},{f^1}, \ldots ,{f^{{2^B} - 1}}} \}$ with $f^n = {\text{e}^{\text{j}\frac{{2\pi }}{{{2^B}}}n}}$ is the set of all the possible quantized phase shifts and $B$ is the number of quantization bits. Then, based on the previous derivation (\ref{eqn:EI-term-theta}) in the WMMSE-EI algorithm, the contribution of $\mathbf{f}_q(p)$ in the objective function can also be obtained. Thus, with the constraint of the quantized phase shifts, the optimization of $\mathbf{f}_q( {p})= {\text{e}^{\text{j}{\theta _{pq}}}}$ becomes 
\begin{equation}
    \theta _{pq}^{{\text{opt}}} = \mathop {\arg \max }\limits_{{\text{e}^{j{\theta _{pq}}}} \in \mathcal{F}} \sum\limits_{k = 1}^K {\frac{{{A_k} + {B_k}\cos \left( {{\theta _{pq}} + {\theta _1}_k} \right)}}{{{C_k} + {D_k}\cos \left( {{\theta _{pq}} + {\theta _2}_k} \right)}}}.
\end{equation}
When the number of quantization bits is small, for example, $B=1$ or $B=2$, the optimal phase can be quickly obtained through a few comparisons. We refer to this modification of WMMSE-EI as the WMMSE-EI-Q algorithm.



Similarly, in the MMSE-EI algorithm with finite phase resolution, the precoding optimization problem becomes the one to select the best phase shift in the $\mathcal{F}$ that maximizes (\ref{eqn:MMSE-objective-EI-max}), i.e., the closest quantized phase shift to $\theta_1$. We refer to this modification of MMSE-EI as the MMSE-EI-Q algorithm.


\section{System Evaluation} \label{sec:analysis}
In this section, we first show the convergence of the proposed WMMSE-EI and WMMSE-MO HBF optimization algorithms. We also analyze and compare the computational complexity of different HBF algorithms.

\subsection{Convergence Analysis} \label{subsec:convergence}
We have shown in Section \ref{sec:design} that the proposed WMMSE-EI and WMMSE-MO HBF optimization algorithms have two levels of iterations. One is the iteration between the hybrid precoding and the hybrid combining optimization with the alternating minimization, which we refer to as the outer iteration, the other is the iteration within the EI or MO algorithm when performing the analog precoding or combining optimization, which we refer to as the inner iteration. Furthermore, we have mentioned in Section \ref{subsec:alternating} that each outer iteration of both the WMMSE-EI and WMMSE-MO algorithms consists of three steps, i.e., hybrid precoding optimization, hybrid combining optimization, weight matrix optimization in series. 

In this subsection, we first show that both the WMMSE-EI and WMMSE-MO algorithms lead to a non-increasing sequence of the matrix weight sum-MSE until the stop criterion is satisfied. Then, from the relationship between the WMMSE problem and the SEM problem, we show that these two algorithms lead to a non-decreasing sequence of the spectral efficiency. 


\begin{proposition}\label{wmmse-pro}
Define the resulting matrix weighted sum-MSE (i.e., the objective function of (\ref{prb:wmmse-broadband})) of the three steps in the $n$-th outer iteration of the WMMSE-EI or the WMMSE-MO algorithms as $J_{s1}^{(n)}$, $J_{s2}^{(n)}$, and $J_{s3}^{(n)}$, respectively. Then, $\{J_{s1}^{(n)}, J_{s2}^{(n)}, J_{s3}^{(n)}\}$ for $n=0,1,\ldots,$ is a non-increasing sequence until the stop condition is satisfied. That is,
\begin{equation}\label{eqn:J-non-increasing}
J_{s1}^{(n)}\geq  J_{s2}^{(n)}\geq J_{s3}^{(n)}\geq
J_{s1}^{(n+1)}\geq  J_{s2}^{(n+1)}\geq J_{s3}^{(n+1)}.
\end{equation}
\end{proposition}

\textit{Proof}: See Appendix B. \hfill $\blacksquare$  




\begin{proposition}\label{rate-pro}
Define the resulting spectral efficiency in the $n$-th outer iteration of the WMMSE-EI or the WMMSE-MO algorithms as $R^{(n)}$. Then, $\{R^{(n)}\}$ for $n=0,1,\ldots,$ is a non-decreasing sequence until the stop condition is satisfied. That is, $R^{(n)}\leq R^{(n+1)}$.
\end{proposition}

\textit{Proof}: See Appendix C. \hfill $\blacksquare$  

\noindent \textit{Remark 1}:
The above convergence proof is based on the optimization order of the hybrid precoder, the hybrid combiner, and the weight matrix within each outer iteration. As we mentioned in Section \ref{subsec:alternating}, the outer iteration can also start with the optimization of hybrid combiner. However, once the hybrid combiner is optimized, the weight matrix should be updated immediately before the optimization of the precoder to ensure the convergence. This can be explained as follows. According to Theorem \ref{theorem:rate-wmmse-eq}, the equivalence between the WMMSE problem and the SEM problem is established using the optimal weight matrix, which is obtained by substituting the optimized  ${\mathbf{W}}_{{\text{D}},k}^{{\text{mmse}}}$ into the MSE matrix. In particular, 
considering the $n$-th outer iteration and assuming that the iteration starts with the hybrid combiner optimization, the  optimized hybrid combiner is denoted by $\mathbf{W}_{k}^{(n)}=\mathbf{W}_{{\text{RF}}}^{(n)}\mathbf{W}_{{\text{D}},k}^{(n)}$, which is a function of the hybrid precoder in the $(n-1)$-th iteration, i.e., ${\mathbf{F}}_k^{(n-1)}$. By then substituting the optimized hybrid combiner into the MSE matrix and the weight matrix, the current spectral efficiency can be expressed as

\begin{equation}\label{eqn:R-remark1}
\begin{aligned}
{R^{\left( n \right)}} &= {\frac{1}{K}\sum\limits_{k = 1}^K {\log \left| {{{\left( {{\bf{E}}_k^{(n)}} \right)}^{ - 1}}} \right|} }\\
&=\frac{1}{K}\sum\limits_{k = 1}^K     \log \left| {{{\mathbf{I}}_{{N_{\text{s}}}}} + \frac{{N_{\text{r}}^{{\text{RF}}}}}{{{\sigma ^2}M}}{{\left( {{\mathbf{W}}_{{\text{RF}}}^{(n)}} \right)}^H}{{\mathbf{H}}_k}{\mathbf{F}}_k^{(n - 1)}{{\left( {{\mathbf{F}}_k^{(n - 1)}} \right)}^H}{\mathbf{H}}_k^H{\mathbf{W}}_{{\text{RF}}}^{\left( n \right)}} \right|.
\end{aligned}
\end{equation}
The convergence of $R^{(n)}$ can be proved in the same procedure as that in Proposition \ref{rate-pro}. Note that $R^{(n)}$ in (\ref{eqn:R-remark1}) is a function of ${\bf{F}}_k^{(n-1)}$. The optimized precoder ${\bf{F}}_k^{(n)}$ in the $n$-th iteration will be used in optimization of combiner and weight matrix in the next iteration.


\noindent \textit{Remark 2}:
So far, we have mainly focused on the HBF design in the case when $N_{\text{r}}^{{\text{RF}}} = {N_{\text{s}}}$, and shown that the SEM problem is equivalent to  the WMMSE problem from Theorem \ref{theorem:rate-wmmse-eq}. We have also proposed the WMMSE-EI and WMMSE-MO logarithms to solve the WMMSE problem with guaranteed convergence. In the case when $N_{\text{r}}^{{\text{RF}}} > {N_{\text{s}}}$, from the proof of Theorem \ref{theorem:rate-wmmse-eq} and (\ref{eqn:WD-rate-MMSE}), we can see that the WMMSE problem is equivalent to the SEM problem maximization if we define the spectral efficiency as the one at the output of the analog combiner $\bf{W}_{\text{RF}}$. In this case, if a linear digital combiner is employed at each subcarrier after the analog combining, there will be always rate reduction after the digital combining. According to \cite{TSE:2005}, an MMSE estimator along with successive interference cancellation can compensate for such rate reduction. Nevertheless, the linear digital combiner obtained in the WMMSE-EI and WMMSE-MO algorithms, which is given by (\ref{eqn:WD-opt}), can still be used. Simulation results in Section \ref{subsec:sim-spe-SNR} will show that the proposed WMMSE based HBF design approach can significantly outperform the conventional one in both the case of $N_{\text{r}}^{{\text{RF}}} > {N_{\text{s}}}$ and the case of $N_{\text{r}}^{{\text{RF}}} = {N_{\text{s}}}$. 

\begin{table}[]
\centering
	{ \caption{Computational complexity of different HBF algorithms}}
	{
\begin{tabular}{|c|c|c|c|c|}
\hline
Proposed Algorithms & Computational Complexity                                               & ${N_{{\text{in}}}}$ &  ${N_{{\text{out}}}}$ & $N_g$ \\ \hline
The WMMSE-EI algorithm & \begin{tabular}[c]{@{}l@{}}${N_\text{out}}{N_{{\text{in}}}}{N_{{\text{ant}}}}K(2N_{{\text{ant}}}^2{N_{{\text{RF}}}} + 3{N_{{\text{ant}}}}N_{{\text{RF}}}^{\text{2}} + 4N_{{\text{ant}}}^2 + 2{N_{{\text{ant}}}}$\\ $\quad \quad  +3N_{{\text{RF}}}^3 - N_{{\text{RF}}}^{\text{2}} - {N_{{\text{ant}}}}{N_{{\text{RF}}}} + {N_g} + 2\mathcal{O}(N_{{\text{RF}}}^3))$\end{tabular} & 3 &10  & 8.1 \\ \hline
 The WMMSE-MO algorithm & \begin{tabular}[c]{@{}l@{}}${N_\text{out}}{N_{{\text{in}}}}(K(5N_{{\text{ant}}}^2{N_{{\text{RF}}}} + 6{N_{{\text{ant}}}}N_{{\text{RF}}}^2 + 4N_{{\text{RF}}}^3 + 4\mathcal{O}(N_{{\text{RF}}}^3))$\\ $ + 3{N_{{\text{ant}}}}{N_{{\text{RF}}}}+{{N_{{\text{ant}}}}}) $\end{tabular} & 21.2 & 10 &  \diagbox{}{}  \\ \hline
The MMSE-EI algorithm  & ${N_\text{out}}{N_{{\text{in}}}}(K(2N_{{\text{ant}}}^2{N_{{\text{RF}}}} + 3{N_{{\text{ant}}}}N_{{\text{RF}}}^{\text{2}} + N_{{\text{RF}}}^3 + \mathcal{O}(N_{{\text{RF}}}^3)) + N_{{\text{ant}}}^2)$                                              &4 &5.2&\diagbox{}{}  \\ \hline
\end{tabular}
}
\end{table}

\subsection{Complexity Analysis}\label{subsec:complexity-ana}
In this subsection we analyze and compare the computational complexity of different HBF optimization algorithms in terms of the number of complex multiplications. As the optimized digital beamformers and the weight matrix have closed-form expressions and have much lower dimension than those of the analog beamformers, we ignore their computational complexity. Besides, as both the analog precoder and combiner can be solved in the same procedure, we focus on the complexity analysis of the analog precoder. To simplify the notation, denote ${N_{{\text{ant}}}} = \max \left\{ {M,N} \right\}$, $N_\text{out}$ as the number of the outer iterations and $N_\text{in}$ as the number of the inner iterations and assume $N_{\text{t}}^{{\text{RF}}} = N_{\text{r}}^{{\text{RF}}} = {N_{\text{s}}} = {N_{{\text{RF}}}}$.


\subsubsection{The WMMSE-EI Algorithm}\label{subsubsec:complexi-analysis-EI}

The complexity of the WMMSE-EI algorithm mainly includes the following two parts:
\begin{itemize}
\item Computation of some related parameters: According to (\ref{fFRF}), the complexity for computing ${{\boldsymbol{\Omega}}_{q,k}}$, ${{\mathbf{A}}_{q,k}}$ and ${{\mathbf{B}}_{q,k}}$ is
$K(2N_{{\text{ant}}}^2{N_{{\text{RF}}}} + 3{N_{{\text{ant}}}}N_{{\text{RF}}}^{\text{2}} + 3N_{{\text{RF}}}^3 - N_{{\text{RF}}}^{\text{2}} - {N_{{\text{ant}}}}{N_{{\text{RF}}}} + 2\mathcal{O}(N_{{\text{RF}}}^3))$, where ${2\mathcal{O}\left( {N_{{\text{RF}}}^3} \right)}$ results from the inversion of two ${N_{{\text{RF}}}} \times {N_{{\text{RF}}}}$ matrices. Furthermore, the complexity for 
computing ${A_k}$, ${B_k}$, ${C_k}$, ${D_k}$, ${\theta _{1,k}}$ and ${\theta _{2,k}}$ in (\ref{eqn:EI-ABCD}) is $ K({4N_{{\text{ant}}}^2 + 2{N_{{\text{ant}}}}})$.

\item One-dimensional line search: We take the golden section search as an example. The complexity is $K{N_g}$ if ${N_g}$ iterations are needed.
\end{itemize}

Thus, the total complexity of the analog precoder optimization using Algorithm \ref{alg:1} is given by
\begin{equation}
\begin{aligned}
{C_{{\text{WMMSE-EI}}}} &={N_\text{out}}{N_{{\text{in}}}}{N_{{\text{ant}}}}K(2N_{{\text{ant}}}^2{N_{{\text{RF}}}} + 3{N_{{\text{ant}}}}N_{{\text{RF}}}^{\text{2}} + 4N_{{\text{ant}}}^2  \\&  + 2{N_{{\text{ant}}}} + 3N_{{\text{RF}}}^3 - N_{{\text{RF}}}^{\text{2}} - {N_{{\text{ant}}}}{N_{{\text{RF}}}} + {N_g} + 2\mathcal{O}(N_{{\text{RF}}}^3)).
\end{aligned}
\end{equation} 

\subsubsection{The WMMSE-MO Algorithm}\label{eqn:complexity-MO} 
The complexity of the WMMSE-MO algorithm mainly includes the following three parts:
\begin{itemize}

\item Computation of the conjugate gradient: From the expression of the conjugate gradient in (\ref{eqn:CG}), the computational complexity is 
$K\left( 4{N_{{\text{ant}}}^2{N_{{\text{RF}}}} + 4{N_{{\text{ant}}}}N_{{\text{RF}}}^2 + 3N_{{\text{RF}}}^3 + 2\mathcal{O}\left( {N_{{\text{RF}}}^3} \right)} \right) + {N_{{\text{ant}}}}{N_{{\text{RF}}}}$,
where ${2O\left( {N_{{\text{RF}}}^3} \right)}$ results from the inversion of ${{\boldsymbol{\Lambda}}_k}$ and ${{\mathbf{M}}_k}$, and the last term ${N_{{\text{ant}}}}{N_{{\text{RF}}}}$ comes from the operation of the Hadamard production in (\ref{pca-gra}).

\item Orthogonal projection and retraction operations: In the MO method, the orthogonal projection and retraction operations are the key steps which map the Euclidean gradient into the Riemannian manifold and guarantee the satisfaction of the constant modulus constraint. According to \cite{Yu:2016}, the complexity of the orthogonal projection and  retraction operations is $2{N_{{\text{ant}}}}{N_{{\text{RF}}}}$ and ${N_{{\text{ant}}}}$, respectively.

\item Armijo backtracking line search: In order to determine a suitable step size, the main complexity of the well-known Armijo backtracking line search is 
$K(N_{{\text{ant}}}^2{N_{{\text{RF}}}} + 2{N_{{\text{ant}}}}N_{{\text{RF}}}^2 + N_{{\text{RF}}}^3 + 2\mathcal{O}(N_{{\text{RF}}}^3))$ matrices.

\end{itemize}

Thus, the total complexity of the analog precoder design using Algorithm \ref{alg:2} is given by
\begin{equation}
{C_{{\text{WMMSE-MO}}}}= {N_\text{out}}{N_{{\text{in}}}}(K(5N_{{\text{ant}}}^2{N_{{\text{RF}}}} + 6{N_{{\text{ant}}}}N_{{\text{RF}}}^2 + 4N_{{\text{RF}}}^3 + 4\mathcal{O}(N_{{\text{RF}}}^3)) + 3{N_{{\text{ant}}}}{N_{{\text{RF}}}}+{{N_{{\text{ant}}}}}).
\end{equation}

\subsubsection{The MMSE-EI Algorithm}\label{subsubsec:complexity-mmse-ei}
The complexity of the MMSE-EI algorithm proposed in Section~\ref{subsec:MMSE} mainly includes the following two parts:
\begin{itemize}
\item Computation of some related parameters: In each inner iteration, the complexity for computing ${\mathbf{A}}$ in (\ref{equ:mmse-UB}) is $K\left( {2N_{{\text{ant}}}^2{N_{{\text{RF}}}} + 3{N_{{\text{ant}}}}N_{{\text{RF}}}^{\text{2}}+{N_{{\text{RF}}}^3} + \mathcal{O}\left( {N_{{\text{RF}}}^3} \right)} \right)$, where ${\mathcal{O}\left( {N_{{\text{RF}}}^3} \right)}$ results from the inversion of an ${N_{{\text{RF}}}} \times {N_{{\text{RF}}}}$ matrix.

\item Update of the optimal phase: The complexity of computing ${\theta _1}$ is ${{N_{{\text{ant}}}}}$. Hence, the complexity of optimizing ${{{\mathbf{F}}_{{\text{RF}}}}}$ is $N_{{\text{ant}}}^2$.

\end{itemize}

Thus, the total complexity of the analog precoder optimization is given by
\begin{equation}
{C_{{\text{MMSE-EI}}}} = {N_\text{out}}{N_{{\text{in}}}}(K(2N_{{\text{ant}}}^2{N_{{\text{RF}}}} + 3{N_{{\text{ant}}}}N_{{\text{RF}}}^{\text{2}} + N_{{\text{RF}}}^3 + \mathcal{O}(N_{{\text{RF}}}^3)) + N_{{\text{ant}}}^2).  
\end{equation}

In summary, the complexity of all the three proposed HBF algorithms is listed in Table I. It can be seen from this table that the dominant term for the three algorithms is ${2N_\text{out}}{N_{{\text{in}}}}{N^3_{{\text{ant}}}}K{N_{{\text{RF}}}}$, $5N_\text{out}{N_{{\text{in}}}}{N^2_{{\text{ant}}}}K{N_{{\text{RF}}}}$, and $2N_\text{out}{N_{{\text{in}}}}{N^2_{{\text{ant}}}}K{N_{{\text{RF}}}}$, respectively, which imply that $C_{\text{WMMSE-EI}}>C_{\text{WMMSE-MO}}\\>C_{\text{MMSE-EI}}$. For more precise comparison and for a more intuitive expression, the average numbers of iterations in different levels are provided over 100 independent channel realizations in simulations, where ${N_{{\text{ant}}}} = 32$, ${N_{{\text{RF}}}} = 4$ and $K=64$. Using these parameters, the number of complex multiplications required by the three algorithms are about $9.0\times 10^8$, $3.3\times 10^8$ and $1.3\times 10^7$, respectively. Thus, the MMSE-EI algorithm has the lowest complexity, with at least one order of magnitude lower than that of the other two algorithms.


\section{Simulation Results}\label{sec:simulation}
In this section, we first present some  simulation results to evaluate the convergence of the proposed HBF optimization algorithms. We then compare the spectral efficiency performance of different HBF algorithms for various system configurations. Finally, we present some results with the consideration of finite resolution phase shifters. 

Consider an mmWave MIMO-OFDM systems with the partially-connected HBF architecture as that in Fig. \ref{fig:system-model}. Unless otherwise specified, we assume that the transmitter takes a half-wavelength spaced ULA with $N=64$ antennas and $N_{\text{t}}^{\text{RF}}=4$ RF chains for the transmission of $N_\text{s}=2$ streams, and the receiver takes a ULA with $M=32$ antennas and $N_{\text{r}}^{\text{RF}}=2$ RF chains. The total number of subcarriers is set to $K=64$. The MIMO channel is generated according to the model in (\ref{eqn:channel-model}) in Section \ref{sec:sysMod-proFormu}, where the number of clusters and the number of rays in each cluster are set to ${N_{\text{C}}} = 5$ and ${N_{\text{R}}} = 10$, respectively, as similar to that in \cite{Sohrabi:2017,Lin:2019}. The complex gain of each ray is assumed to satisfy the circularly symmetric complex Gaussian distribution with ${{{h_{cl}}}} \sim \mathcal{C}\mathcal{N}\left( {0,1} \right)$. The AoA, $\theta _{cr}^{\text{r}}$, and AoD, $\theta _{cr}^{\text{t}}$, are generated according to the Laplacian distribution with random mean cluster angles $\bar \theta _{cr}^{\text{r}} \in [0,2\pi )$ and $\bar \theta _{cr}^{\text{t}} \in [0,2\pi )$. 

\subsection{Convergence Behavior}\label{subsec:sim-convergence}
\begin{figure}
 	\begin{minipage}[c]{0.45\linewidth}
 		\centering
 		\includegraphics[width=3.1in]{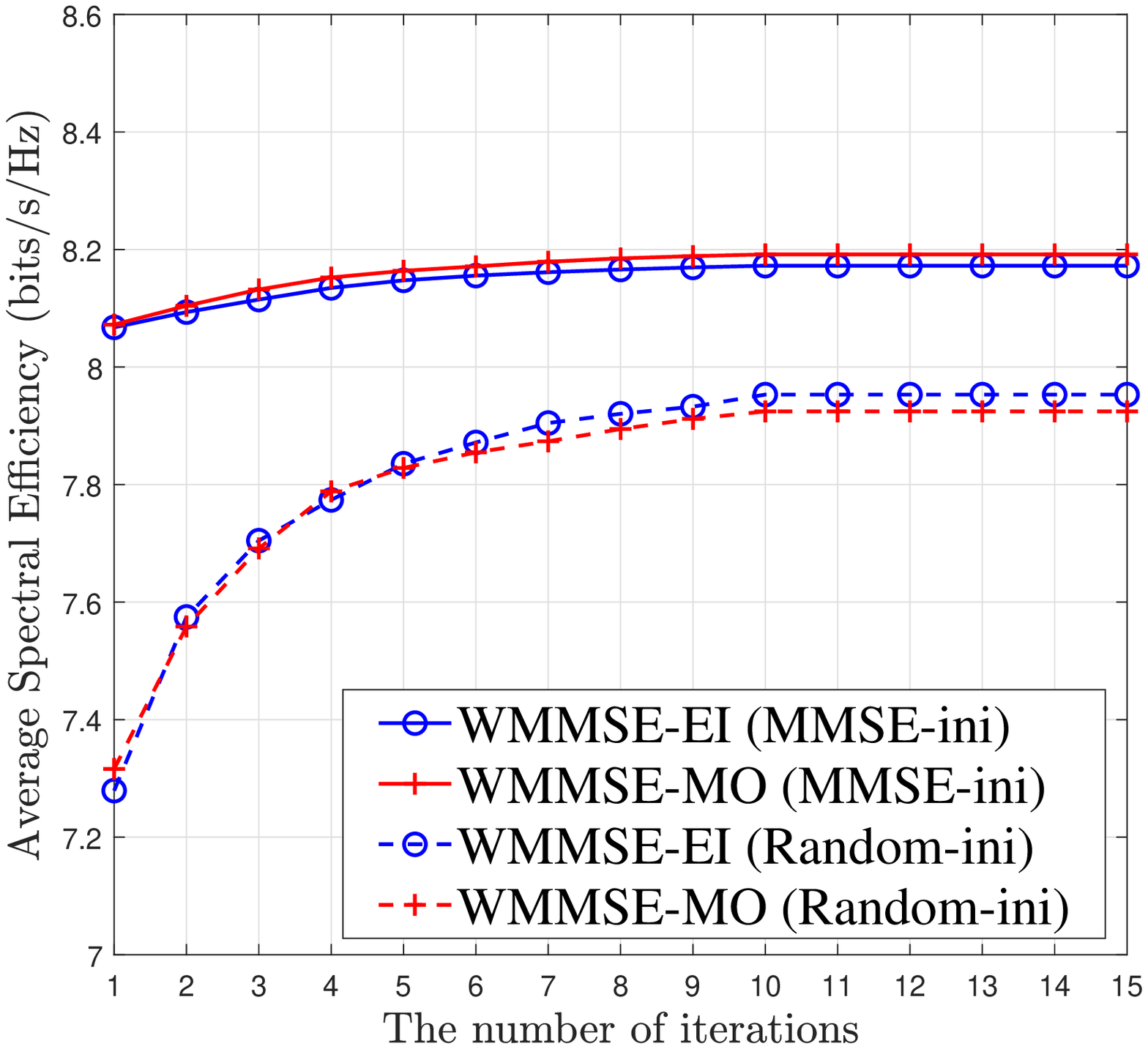}
 		\caption{Convergence properties of the WMMSE-EI and WMMSE-MO HBF algorithms with different  initialization methods.}
 		\label{Fig1}
 				\vspace{-0.2cm}
 	\end{minipage}%
 	\hspace{1.4cm}
 	\begin{minipage}[c]{0.45\linewidth}
 		\centering
 		\includegraphics[width=3.1in]{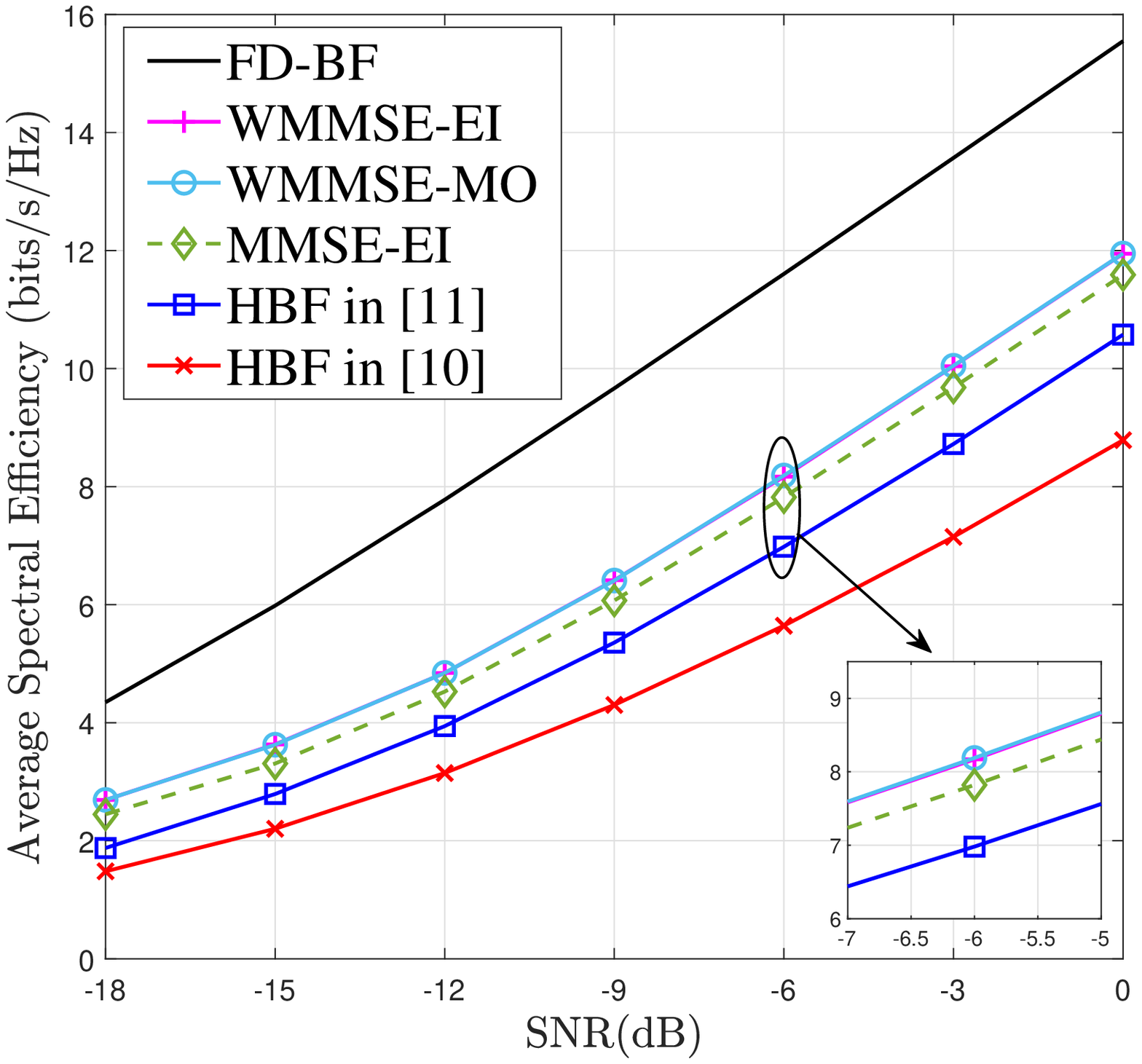}
 		\caption{Spectral efficiency v.s. SNR for different HBF algorithms for a $64 \times 32$ MIMO-OFDM system with  $N_{\text{t}}^{\text{RF}}=4,N_{\text{r}}^{\text{RF}}={N_{\text{s}}} = 2$.}
 		\label{Fig2}
 			\vspace{-0.2cm}	
 	\end{minipage}
 \end{figure}

First, we evaluate the convergence properties of the proposed HBF WMMSE-EI and WMMSE-MO algorithms. It is worth noting that the performance of these two iterative algorithms is highly related to the initialization of the beamforming optimization variables. We selected two different initialization methods for testing: one was random initialization (labeled with `Random-ini'), the other was to use the low complexity MMSE-EI algorithm for initialization (labeled with 'MMSE-ini'). Fig. \ref{Fig1} illustrates the performance of average spectral efficiency as a function of the number of outer iterations (the iteration between the hybrid precoding and combining using the alternative minimization) for the proposed WMMSE-EI and WMMSE-MO algorithms with the two initialization methods when SNR is fixed at $-6\text{dB}$. According to Fig. \ref{Fig1}, with either of these two initialization methods, the WMMSE-EI and WMMSE-MO algorithms both converge, which verifies the convergence proof in Section \ref{subsec:convergence}. Furthermore, Fig. \ref{Fig1} also shows that the `MMSE-ini' method can significantly speed up the convergence and further improve the spectral efficiency compared with the random initialization. Thus,  the `MMSE-ini' method is used for initialization in the following simulations.


\subsection{Spectral efficiency v.s. SNR}\label{subsec:sim-spe-SNR}
Fig. \ref{Fig2} shows the performance of spectral efficiency as a function of SNR for the proposed WMMSE-EI, WMMSE-MO and MMSE-EI algorithms. For comparison, the performance of two conventional HBF algorithms for the partially-connected architecture (labeled with `HBF in \cite{Yu:2016}' and `HBF in \cite{Sohrabi:2017}') and that of the optimal fully-digital beamforming (labeled with `FD-BF') are also provided in Fig. \ref{Fig2}. It is shown that the proposed WMMSE-EI and WMMSE-MO HBF algorithms perform almost the same, and significantly outperform the conventional HBF algorithm in \cite{Yu:2016} and that in \cite{Sohrabi:2017} by about more than $4.5\text{dB}$ and $2.0\text{dB}$ in SNR, respectively, for a target spectral efficiency of $8\text{bits/s/Hz}$. Besides, the proposed low complexity MMSE-EI algorithm has a gap of about $0.5\text{dB}$ in SNR when compared to the proposed WMMSE-EI and WMMSE-MO algorithms. Furthermore, it can be seen from this figure that the performance gap between the optimal fully-digital beamforming and the HBF with the partially-connected architecture is relatively large mainly due to the great reduction of the number RF chains and the number of phase shifters in the partially-connected architecture. Nevertheless, the proposed WMMSE based HBF algorithms provide a more promising design approach to balance the performance loss and hardware cost and consumption for the HBF design with the partially-connected architecture.

\begin{figure}
 	\begin{minipage}[c]{0.45\linewidth}
 		\centering
 		\includegraphics[width=3.1in]{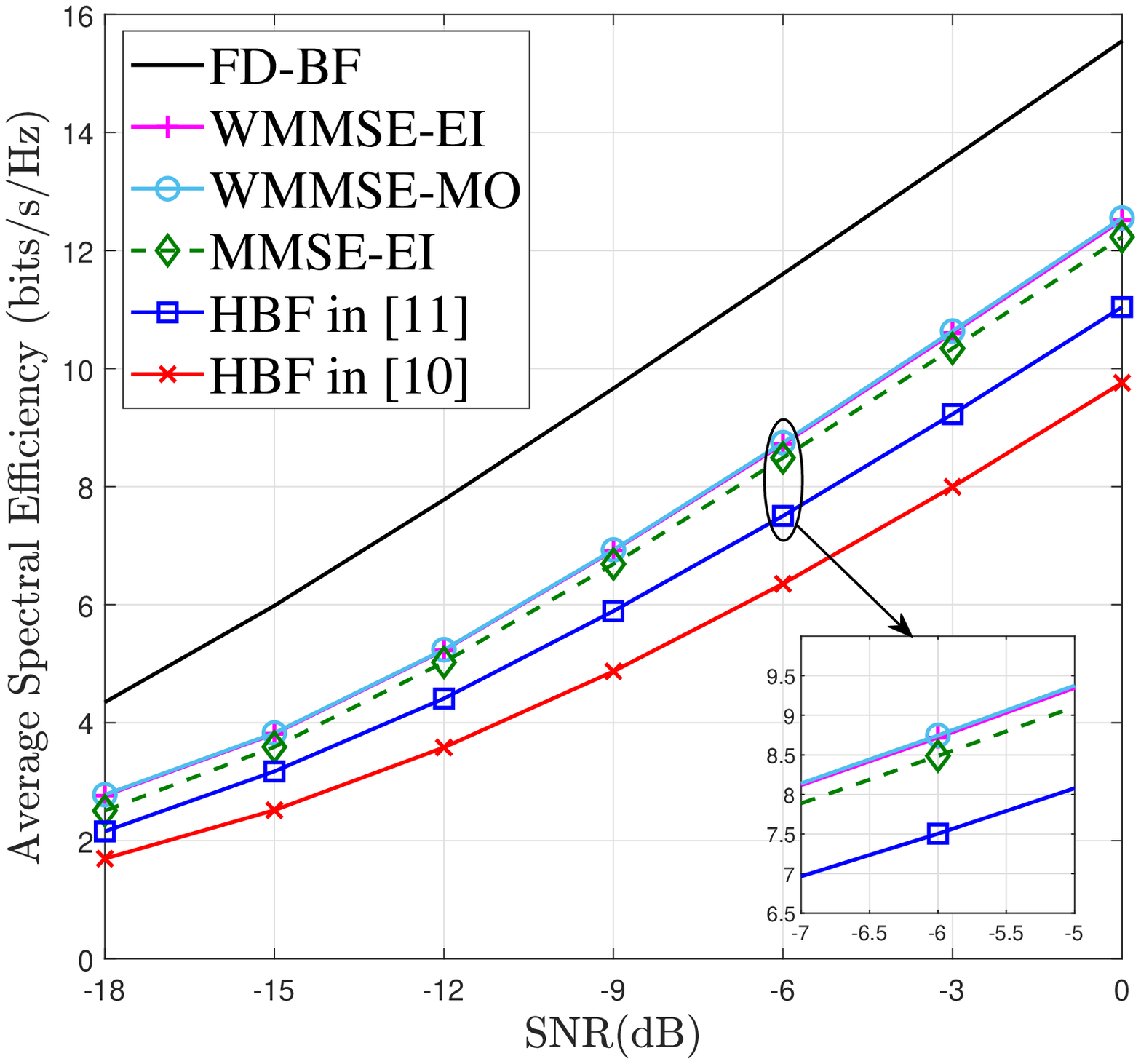}
 		\caption{Spectral efficiency v.s. SNR for different HBF algorithms for a $64 \times 32$ MIMO-OFDM system with  $N_{\text{t}}^{\text{RF}}=N_{\text{r}}^{\text{RF}}=4, {N_{\text{s}}} = 2$.}
 		\label{Fig3}
 				\vspace{-0.2cm}
 	\end{minipage}%
 	\hspace{1.4cm}
 	\begin{minipage}[c]{0.45\linewidth}
 		\centering
 		\includegraphics[width=3.1in]{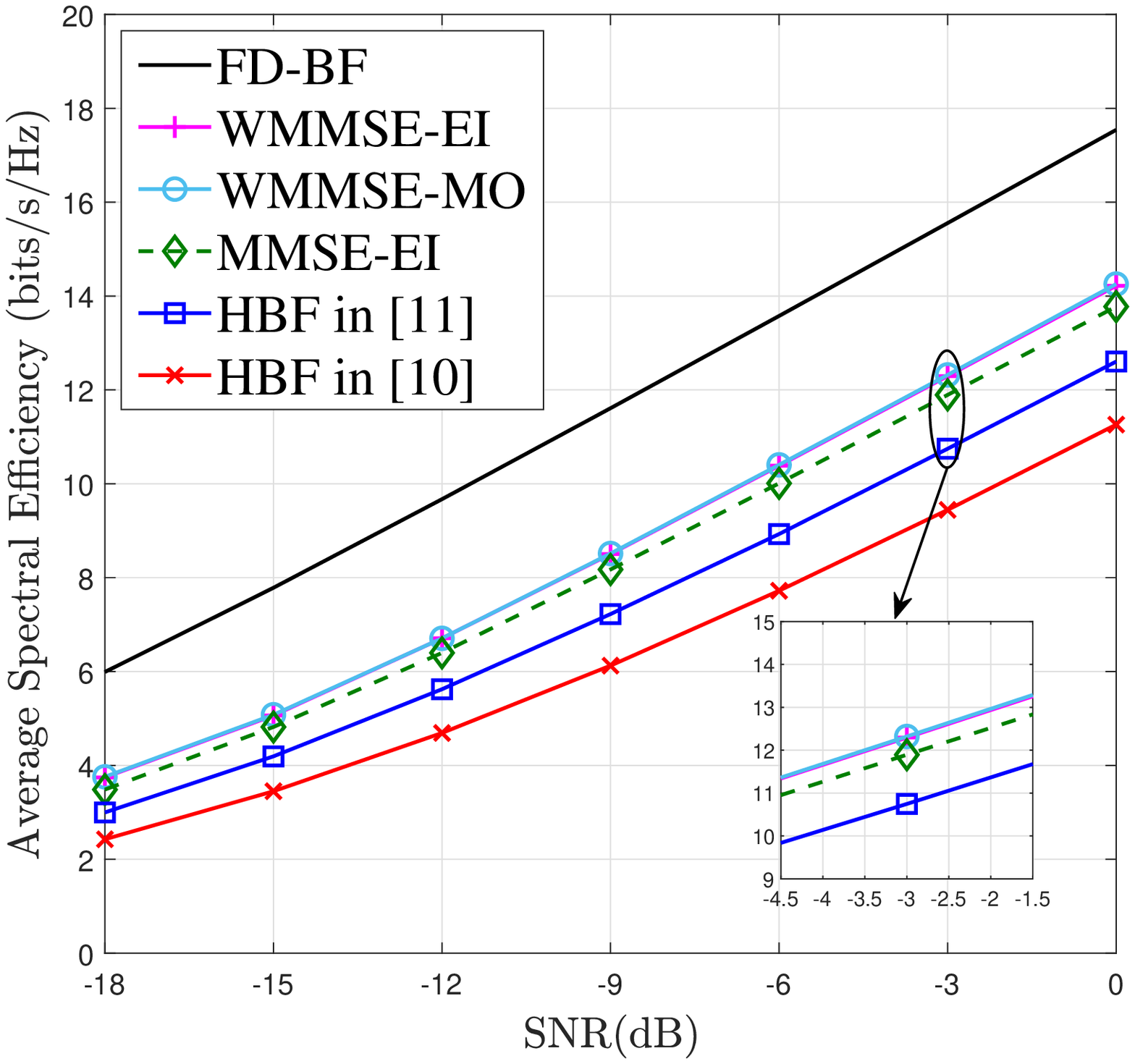}
 		\caption{Spectral efficiency v.s. SNR for different HBF algorithms for a $144 \times 32$ MIMO-OFDM systems with  $N_{\text{t}}^{\text{RF}}=N_{\text{r}}^{\text{RF}}=4, {N_{\text{s}}} = 2$.}
 		\label{Fig4}
 			\vspace{-0.2cm}	
 	\end{minipage}
 \end{figure}

To verify the generality of the proposed HBF algorithms, we consider two other mmWave MIMO system configurations, where more receive RF chains ($N_{\text{r}}^{\text{RF}}=4$) are employed in Fig. \ref{Fig3} and more transmit antennas ($N=144$) are further employed in Fig. \ref{Fig4}. It can be seen from these two figures that the proposed HBF algorithms can also achieve similar performance improvement over the conventional counterparts.  

 
\subsection{Performance with Finite Resolution Phase Shifters} \label{subsec:sim- finite resolution phase shifters}
\begin{figure}
 	\begin{minipage}[c]{0.45\linewidth}
 		\centering
 		\includegraphics[width=3.1in]{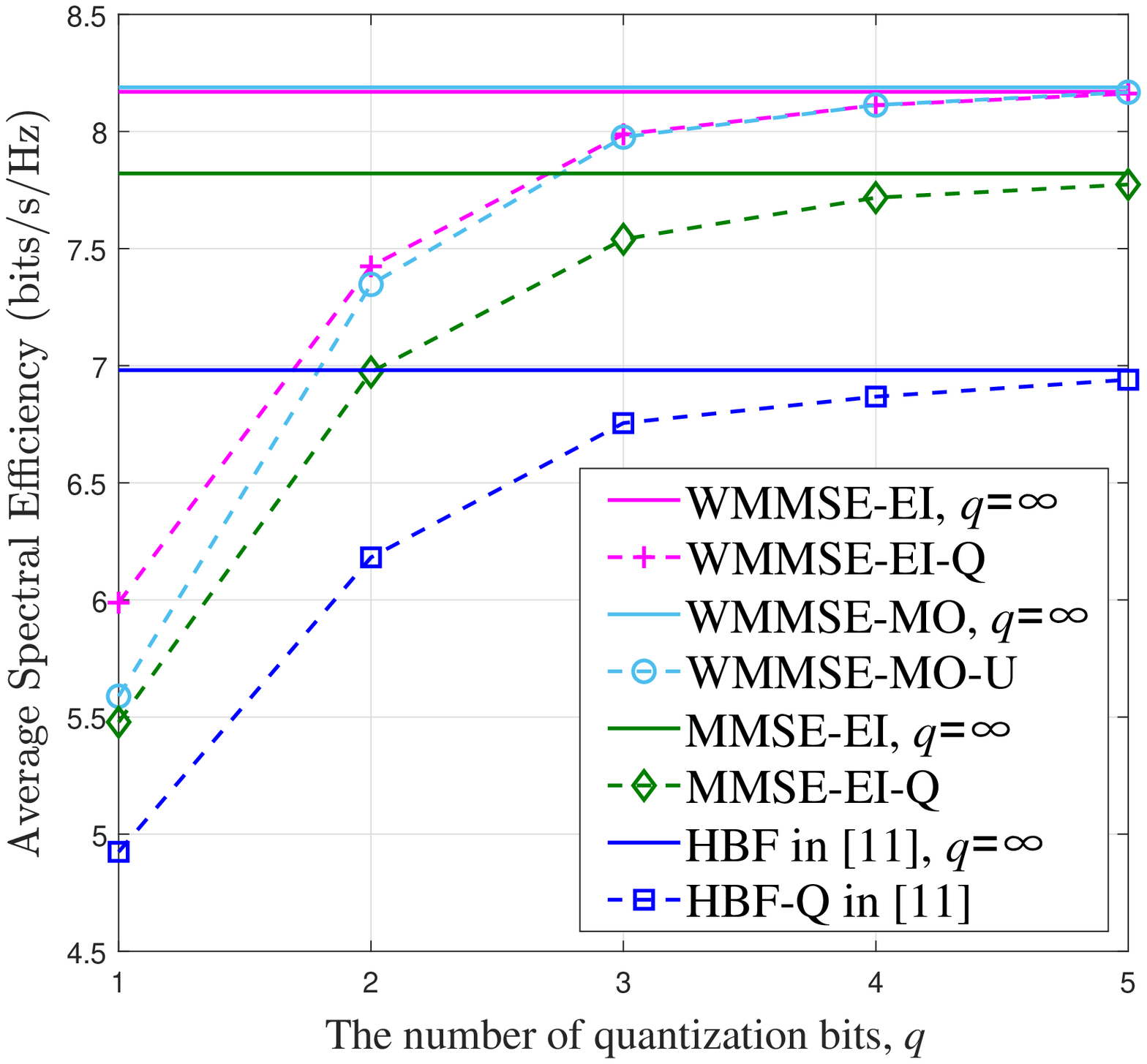}
 		\caption{Spectral efficiency v.s. the number of quantization bits for different HBF algorithms.}
 		\label{Fig5}
 				\vspace{-0.2cm}
 	\end{minipage}%
 	\hspace{1.4cm}
 	\begin{minipage}[c]{0.45\linewidth}
 		\centering
 		\includegraphics[width=3.1in]{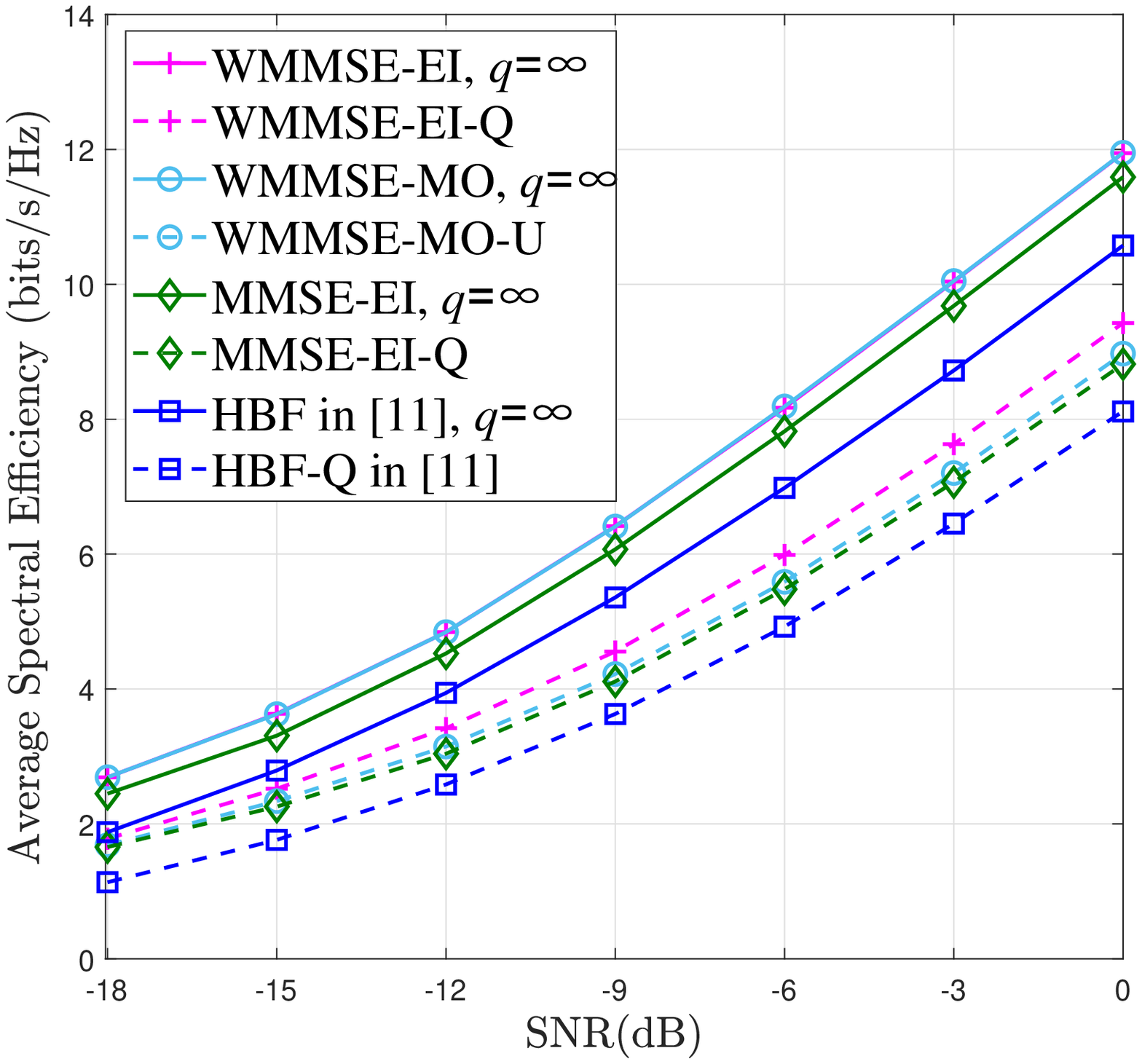}
 		\caption{Spectral efficiency v.s. SNR for different HBF algorithms in the special case of 1-bit phase shifters.}
 		\label{Fig6}
 			\vspace{-0.2cm}	
 	\end{minipage}
 \end{figure}

Considering the fact that practical phase shifters may have limited resolution, we compare the performance of different HBF algorithms with different numbers of quantization bits, denoted by $q$, in Fig. \ref{Fig5} when the SNR is fixed at $-6{\text{dB}}$. For the WMMSE-EI and MMSE-EI algorithms, the modified algorithms with the consideration of finite resolution proposed in Section \ref{subsec:phase-shifter}, i.e., the WMMSE-EI-Q and MMSE-EI-Q algorithms, were applied in the simulation. For the WMMSE-MO algorithm, we first obtained the optimized analog beamforming matrices under the condition of infinite resolution and then simply uniformly quantized the phase of each entry with $q$ bits, which is labeled with `WMMSE-MO-U' in the figure. For comparison, we also provide the performance of the HBF design with finite resolution phase shifters in \cite{Sohrabi:2017}, which is labeled as `HBF-Q in \cite{Sohrabi:2017}' in the figure. It can be seen from this figure that the proposed HBF algorithms still outperform the conventional counterpart with finite phase shift resolution. Meanwhile, the WMMSE-EI-Q algorithm achieves higher spectral efficiency for small $q$ such as $q=1$ or $q=2$ than the uniform quantization method. The MMSE-EI-Q algorithm even outperforms the WMMSE-MO-U algorithm when $q=1$. Fig. \ref{Fig5} also shows that the performance loss caused by finite resolution is almost negligible when $q \geq 4$. 



Finally, Fig. \ref{Fig6} shows the spectral efficiency v.s. SNR for these HBF algorithms in the special case of 1-bit phase shifters. It can be seen from this figure that for all the algorithms with $q=1$, about $3\sim 4\text{dB}$ more power needs to be paid to achieve the same target spectral efficiency with infinite resolution. Furthermore, the performance gain achieved by the proposed algorithms over the conventional algorithms with infinite resolution is maintained and even increased for system with 1-bit phase shifters by using the WMMSE-EI-Q and MMSE-EI-Q algorithms in Section~\ref{subsec:phase-shifter}.  


\section{Conclusions}\label{sec:conclusion}
We have proposed a WMMSE based design approach to equivalently solve the HBF optimization problem with the objective of maximizing the spectral efficiency. To deal with the highly non-convex and multivairate WMMSE problem, we separated it into the hybrid precoding and combining subproblems and applied the alternating minimization algorithm to iteratively optimize the hybrid precoder and combiner. The proposed the WMMSE-EI and WMMSE-MO algorithms have been shown to achieve $2\text{dB}$ SNR gain over the conventional algorithms with guaranteed convergence for various MIMO configurations. Furthermore, we have modified the WMMSE based algorithms to the MMSE based ones to reduce the computational complexity and provide much better initial beamformers to the WMMSE based algorithms than the random initialization. Considering the practical finite resolution phase shifters, we have also proposed some modified HBF algorithms, which perform better than the one using the uniform quantization and the conventional algorithm, especially in the case of 1-bit phase shifters.  

\section*{Appendix A \\ Proof of Lemma 1 }\label{APA}

\textit{Proof}: According to \cite{Zhangxianda}, the direction of the conjugate gradient indicates the direction of curvature for an unconstrained optimization problem. Therefore, the general conjugate gradient of $f({{{\mathbf{F}}_{{\text{RF}}}}})$ respect to ${{{\mathbf{F}}_{{\text{RF}}}}}$ in the Euclidean space without considering the partially-connected architecture is given by
\begin{equation}\label{mo:conj1}
    \frac{{\partial f\left( {{{\mathbf{F}}_{{\text{RF}}}}} \right)}}{{\partial {\mathbf{F}}_{{\text{RF}}}^*}}{\text{ = }}\left[ {\begin{array}{*{20}{c}}
  {\frac{{\partial f\left( {{{\mathbf{F}}_{{\text{RF}}}}} \right)}}{{\partial {\mathbf{F}}_{{\text{RF}}}^*(1,1)}}}&{\frac{{\partial f\left( {{{\mathbf{F}}_{{\text{RF}}}}} \right)}}{{\partial {\mathbf{F}}_{{\text{RF}}}^*(1,2)}}}& \cdots &{\frac{{\partial f\left( {{{\mathbf{F}}_{{\text{RF}}}}} \right)}}{{\partial {\mathbf{F}}_{{\text{RF}}}^*(1,N_{\text{t}}^{{\text{RF}}})}}} \\ 
  {\frac{{\partial f\left( {{{\mathbf{F}}_{{\text{RF}}}}} \right)}}{{\partial {\mathbf{F}}_{{\text{RF}}}^*(2,1)}}}&{\frac{{\partial f\left( {{{\mathbf{F}}_{{\text{RF}}}}} \right)}}{{\partial {\mathbf{F}}_{{\text{RF}}}^*(2,2)}}}& \cdots &{\frac{{\partial f\left( {{{\mathbf{F}}_{{\text{RF}}}}} \right)}}{{\partial {\mathbf{F}}_{{\text{RF}}}^*(2,N_{\text{t}}^{{\text{RF}}})}}} \\ 
   \vdots & \vdots & \ddots & \vdots  \\ 
  {\frac{{\partial f\left( {{{\mathbf{F}}_{{\text{RF}}}}} \right)}}{{\partial {\mathbf{F}}_{{\text{RF}}}^*({N_{\text{t}}},1)}}}&{\frac{{\partial f\left( {{{\mathbf{F}}_{{\text{RF}}}}} \right)}}{{\partial {\mathbf{F}}_{{\text{RF}}}^*({N_{\text{t}}},2)}}}& \cdots &{\frac{{\partial f\left( {{{\mathbf{F}}_{{\text{RF}}}}} \right)}}{{\partial {\mathbf{F}}_{{\text{RF}}}^*({N_{\text{t}}},N_{\text{t}}^{{\text{RF}}})}}} 
\end{array}} \right].
\end{equation}
Now considering the partially-connected architecture, as a RF chain is only connected to part of the antennas, the objective function $f({{{\mathbf{F}}_{{\text{RF}}}}})$ is only related to the  block-diagonal terms in ${{{\mathbf{F}}_{{\text{RF}}}}}$ given by (\ref{eqn:F_RF_block_diag}). Thus, at those positions where the entries of the matrix ${{{\mathbf{F}}_{{\text{RF}}}}}$ are equal to zero (i.e., the non-block-diagonal entries), the corresponding partial derivative entries in (\ref{mo:conj1}) should be zero. Therefore, the Euclidean conjugate gradient $\nabla f\left( {{{\mathbf{F}}_{{\text{RF}}}}} \right)$ for the partially-connected architecture must be a block-diagonal matrix as follow
\begin{equation}\label{mo:conj2}
\nabla f\left( {{{\mathbf{F}}_{{\text{RF}}}}} \right)={\text{blkdiag}}\left( {\frac{{\partial f\left( {{{\mathbf{F}}_{{\text{RF}}}}} \right)}}{{\partial {{\mathbf{f}}_1}}},\frac{{\partial f\left( {{{\mathbf{F}}_{{\text{RF}}}}} \right)}}{{\partial {{\mathbf{f}}_2}}}, \cdots ,\frac{{\partial f\left( {{{\mathbf{F}}_{{\text{RF}}}}} \right)}}{{\partial {{\mathbf{f}}_{N_{\text{t}}^{{\text{RF}}}}}}}} \right)={\nabla _{{\mathbf{F}}_{{\text{RF}}}^*}}f\left( {{{\mathbf{F}}_{{\text{RF}}}}} \right) \odot {{\mathbf{P}}_1}.
\end{equation}
The proof is completed.  \hfill $\blacksquare$

\section*{Appendix B \\ Proof of Proposition 1}\label{APB}
\textit{Proof}: First, as we have shown in (\ref{eqn:optimal-FDk}) and (\ref{eqn:WD-opt}) that the optimal digital precoder (along with the optimal $\xi_k$ in (\ref{eqn:Xi-optimal})) and combiner have a closed-form solution obtained via the KKT conditions, for a given analog precoder or combiner, the corresponding digital one always ensures the decrease of the weighted sum-MSE\cite{Boyd:2004}. Therefore, it is the optimization of analog beamformers that decides the convergence of the HBF algorithms. For the EI algorithm, as it is an element-by-element optimization algorithm, for each inner iteration, the updated phase shifter element always guarantees that the objective function will not increase. For the MO algorithm, as we have shown in Section \ref{subsubsec:MO-algorithm}, the inner iteration converges to a local optimal point. Thus, the processing within the hybrid precoding or hybrid combining optimization always ensures the convergence of the objective function. 

Next, when the optimization switches from the hybrid precoding to the hybrid combining, as the updated hybrid precoder is taken as a fixed one during the whole inner iteration of the hybrid combining optimization, the objective function of the matrix weighted sum-MSE does not increase. This is also similar to the optimization of the weight matrix and that of the hybrid precoding, where all the other optimization variables are fixed except for the one that needs to be optimized in the current step. The proof is thus completed. \hfill $\blacksquare$ 

\section*{Appendix C \\ Proof of Proposition 2}\label{APC}
\textit{Proof}: For the $n$-th outer iteration, after the first and second step, both the hybrid precoder and the combiner have been updated. Then, from (11), we obtain the updated MSE matrix for each subcarrier, which is defined as ${\mathbf{E}}_k^{\left( n \right)}$. According to Theorem \ref{theorem:rate-wmmse-eq}, the optimal weight matrix in the third step of the WMMSE-EI or WMMSE-MO algorithm should be $\boldsymbol{\Lambda}_k^{(n)} = ({\mathbf{E}}_k^{\left( n \right)})^{-1}$, the resulting objective function of (\ref{prb:mmse-broadband}) in the $n$-th iteration is then given by
\begin{equation}\label{eqn:mse-rate}
  {J_{s3}^{\left( n \right)}}={N_{\text{s}}} -  {\frac{1}{K}\sum\limits_{k = 1}^K {\log \left| {{{\left( {{\bf{E}}_k^{(n)}} \right)}^{ - 1}}} \right|} }={{N_{\text{s}}} - {R^{(n)}}}, 
\end{equation}
where the second equality follows from (\ref{eqn:WD-rate-MMSE}) in Theorem 1. According to Proposition \ref{wmmse-pro}, where we have shown $J_{s3}^{(n)}\geq J_{s3}^{(n+1)}$, we now have ${R^{\left( n \right)}} \leq {R^{\left( {n + 1} \right)}}$. Therefore, the proposed WMMSE-EI and WMMSE-MO algorithms, which are designed based on the WMMSE criterion, can indeed make the spectral efficiency monotonously increase until convergence. The proof is thus completed.\hfill $\blacksquare$  


\bibliography{ref}
\bibliographystyle{IEEEtran}	
\end{document}